\documentclass[12pt]{article}
\usepackage{url}
\usepackage{amsfonts}
\usepackage{amsmath,amssymb,color}
\usepackage{graphicx}
\usepackage{enumitem}
\usepackage{multirow}
\usepackage{bm}
\usepackage{natbib}
\usepackage{verbatim}
\usepackage{float}
\usepackage[toc,page]{appendix}
\usepackage[onehalfspacing]{setspace}
\usepackage{setspace}
\usepackage{adjustbox}
\usepackage{rotating}
\usepackage{subcaption}
\usepackage{amsthm}
\usepackage{booktabs,dcolumn,caption}

\usepackage{array}
\newcolumntype{L}[1]{>{\raggedright\let\newline\\\arraybackslash\hspace{0pt}}p{#1}}
\newcolumntype{C}[1]{>{\centering\let\newline\\\arraybackslash\hspace{0pt}}p{#1}}
\newcolumntype{R}[1]{>{\raggedleft\let\newline\\\arraybackslash\hspace{0pt}}p{#1}}

\newcommand{\ttt}{\boldsymbol \theta}

\newcommand{\YY}{\mbox{$\mathbf Y$}}
\newcommand{\yy}{\mbox{$\mathbf y$}}

\newcommand{\xx}{\mathbf x}

\newcommand{\argmax}{\operatornamewithlimits{arg\,max}}
\newcommand{\argmin}{\operatornamewithlimits{arg\,min}}

\title{Item Response Theory -- A Statistical Framework for Educational and Psychological Measurement}
\author{Yunxiao Chen, London School of Economics and Political Science\\
Xiaoou Li, University of Minnesota\\
Jingchen Liu and Zhiliang Ying, Columbia University}
\date{}

\begin{document}
\maketitle

\begin{abstract}
Item response theory (IRT)
has become one of the most popular statistical
models for psychometrics, a field of study concerned with the theory and techniques of psychological measurement. The IRT models are latent factor models tailored to the analysis, interpretation, and prediction of individuals' behaviors in answering a set of measurement items that typically involve categorical response data.
Many important questions of measurement are directly or indirectly
answered through the use of IRT models, including scoring individuals' test performances, validating a test scale, linking two tests, among others. This paper provides a review of item response theory, including its statistical framework and psychometric applications.  We establish connections between item response theory and related topics in statistics, including empirical Bayes, nonparametric methods, matrix completion, regularized estimation, and sequential analysis.
Possible future directions of IRT are discussed from the perspective of statistical learning.
\end{abstract}	

\section{Introduction}\label{Sec:intro}

Item response theory (IRT) models, also referred to as latent trait models, play an important role in educational testing and psychological measurement as well as several other areas of behavioral and cognitive measurement. Specifically, IRT models have been widely
used in the construction, evaluation, and sometimes scoring, of large-scale high-stakes educational tests \citep[e.g.,][]{Birdsall2011,robin2014multistage}.
Most national and international large-scale assessments for monitoring education quality,
such as the Programme for International Student Assessment (PISA) and the Trends in International Mathematics and Science Study (TIMSS), which are of lower-stakes for test-takers, are also analyzed and reported under the IRT framework \citep{rutkowski2013handbook}.
IRT models are a building block of student learning models for intelligent tutoring systems and personalized learning \citep{chen2018recommendation,khajah2014integrating,khajah2014integrating2,tang2018reinforcement,wilson2015knewton}.
They also play an important role in the analysis of health-related quality of life \citep{cella2002item,hays2000item}; specifically, these models have been the central statistical tool for the development of the Patient-Reported Outcomes Measurement Information System (PROMIS),
a state-of-the-art assessment system for self-reported health
that provides a standardized measurement of physical, mental, and social well-being from patients' perspectives \citep{cella2007patient}.
Furthermore, they are crucial to measuring psychological traits in various domains of psychology, including personality and psychopathology \citep{balsis2017item,reise2009item,wirth2007item}.
Besides these applications, IRT models receive wide applications in many other areas, such as political voting, marketing research, among others \citep[e.g.,][]{bafumi2005practical,de2008using}.

IRT models are probabilistic models for individuals' responses to a set of items (e.g., questions), where the responses are typically categorical (e.g., binary/ordinal/nominal).
These models are latent factor models from a statistical perspective, dating back to Spearman's factor model for intelligence \citep{spearman1904general}. Some early developments on IRT include \cite{richardson1936relation}, \cite{ferguson1942item}, \cite{lawley1943xxiii,lawley1944x}, among others.
In the 1960s,
\cite{rasch1960probabilistic} and \cite{lord1968statistical} laid the foundation of IRT as a theory for educational and psychological testing. Specifically, \cite{rasch1960probabilistic}
proposed  what now known as the Rasch model, an IRT model with a very simple form that has important philosophical implications on psychological measurement and possesses good statistical properties brought by its natural exponential family form.
\cite{lord1968statistical} first introduced a general framework of IRT models  and proposed several parametric forms of IRT models.
In particular, the two-parameter (2PL) and three-parameter logistic (3PL) models \citep{birnbaum1968some} were introduced that are still widely used in  educational testing these days. 
Following these pioneer works, more flexible models and more powerful statistical tools have been developed to better measure human behaviors, promoting IRT to become one of the dominant paradigms for measurement in education, psychology, and related problems; see \cite{carlson2017item}, \cite{embretson2000item},  and \cite{van2018handbook} for the history of IRT.

IRT models are closely related to linear factor models \citep[see e.g.,][]{anderson1956statistical,bai2012statistical}. The major difference is that linear factor models assume that the observed variables are continuous, while  IRT models mainly focus on categorical variables. Due to their close connections, one can view IRT models as factor models for categorical data \citep{bartholomew1980factor}.
IRT models are also similar to
generalized linear mixed models \citep[GLMM;][]{berridge2011multivariate,searle2001generalized} in terms of the model assumptions, even though the two modeling frameworks are developed independently and focus on different inference problems. 
These connections will be further discussed in Section~\ref{sec:model} below.

Classical test theory (CTT; \citealp{lord1968statistical, novick1966axioms,spearman1904proof}), also known as the true score theory, is another psychometric paradigm that was dominant before the prevalence of IRT.
Perhaps the key advantage of IRT over CTT is that
IRT takes item-level data as input and
separately models and estimates the person and item parameters.
This advantage of IRT allows for tailoring tests through judicious item selection to achieve precise measurement for individual test-takers (e.g., in computerized adaptive testing) or designing parallel test forms with the same psychometric properties. It also provides mechanisms for placing different test forms on the same scale (linking and equating) and defining and evaluating test reliability and validity. These features of IRT will be discussed in the rest of the paper.


This paper reviews item response theory, focusing on its statistical framework and psychometric applications.
In the review, we establish connections between IRT and related topics in statistics, such as {empirical Bayes, nonparametric methods, matrix completion, regularized estimation, and sequential analysis.}
The purpose is three-fold: (1) to provide a summary of the statistical foundation of IRT, (2) to introduce some of the major problems in psychometrics to general statistical audiences through IRT as a lens, and (3) to suggest directions of methodological development for solving new measurement challenges in the big data era.


The rest of the paper is organized as follows. In Section~\ref{sec:model}, we provide a review of the statistical modeling framework of IRT and compare it with classical test theory and several related models. In Section~\ref{sec:admin}, we discuss the statistical analyses under the IRT framework and their
psychometric applications. In Section~\ref{sec:future}, we discuss several future directions of methodological development for solving new measurement challenges in the big data era. We end with a discussion in Section~\ref{sec:dis}.

\section{Item Response Theory: A Review and Related Statistical Models}\label{sec:model}

\subsection{Basic Model Assumptions}\label{subsec:model}


To be concrete, we discuss the modelling framework in the context of  educational testing, though it is applicable to many other areas. We have $N$ individuals responding to $J$ test items.
Let $Y_{ij} \in \{0, 1\}$ be a random variable representing test-taker $i$'s response to item $j$, where $Y_{ij} = 1$ indicates a correct response and $Y_{ij} = 0$ otherwise. We further denote $y_{ij}$ as a realization of $Y_{ij}$ and denote $\YY_i = (Y_{i1}, ..., Y_{iJ})^\top$ and $\yy_i = (y_{i1}, ..., y_{iJ})^\top$. Note that
we use boldface for vectors.
An IRT model specifies the joint distribution of $\YY_i, i = 1, ..., N$.


An IRT model assumes that
$\YY_{i}$, $i = 1, ..., N$,
 are independent, and models the joint distribution of random vector $\YY_i$ through the introduction of individual-specific latency. More precisely, a unidimensional IRT model assumes one latent variable for each individual $i$, denoted by $\theta_i$, which is interpreted as the individual's level on a certain latent trait (i.e., ability) measured by the test. It is assumed that individuals' response patterns are completely characterized by their latent trait levels. This is reflected by the conditional distribution of $\YY_i$ given $\theta_i$ in an IRT model. The specification of this conditional distribution relies on two assumptions (1) a \textit{local independence} assumption, saying that $Y_{i1}$, ..., $Y_{iJ}$ are conditionally independent given the latent trait level $\theta_i$, and (2) an assumption on the item response function (IRF), also known as the item characteristic curve (ICC),  defined as $g_j(\theta\vert \boldsymbol\pi_j) := P(Y_{ij} = 1\vert \theta_i = \theta)$, where $\boldsymbol\pi_j$ is a generic notation for parameters of item $j$.  For example, $g_j(\theta\vert \boldsymbol\pi_j)$ takes the form
\begin{equation}\label{eq:rasch}
\frac{\exp(\theta - b_j)}{1+\exp(\theta - b_j)},
\end{equation}
\begin{equation}\label{eq:2pl}
\frac{\exp(d_j + a_j \theta)}{1+\exp(d_j + a_j \theta)},
\end{equation}
and
\begin{equation}\label{eq:3pl}
c_j + (1-c_j)\frac{\exp(d_j + a_j \theta)}{1+\exp(d_j + a_j \theta)},
\end{equation}
in the Rasch,  2PL, and 3PL models \citep{birnbaum1968some,rasch1960probabilistic}, respectively, where item parameters $\boldsymbol \pi_j = b_j$, $(a_j, d_j)$, and $(a_j, c_j, d_j)$, respectively.
Probit (normal ogive) models, which were first proposed in \cite{lawley1943xxiii,lawley1944x},
are also commonly used \citep{embretson2000item}.
These models replace the logit link in \eqref{eq:rasch}-\eqref{eq:3pl} by a probit link. That is, for example, the two-parameter probit model has the IRF
$\Phi(d_j + a_j \theta)$,
where $\Phi$ denotes the standard normal distribution. We also note that different items can have different IRFs.
In summary, a unidimensional IRT model assumes that $Y_{ij}$ is $g_j(\theta_i\vert \boldsymbol \pi_j)$ plus some noise, where the noise is often known as the measurement error. In educational testing, the IRF is usually assumed to be a monotonically increasing function (e.g., $a_j >0$ in the 2PL and 3PL models) so that a higher latent trait level leads to a higher chance of correctly answering the item.

The complete specification of an IRT model remains to impose assumptions on the $\theta_i$.
According to \cite{holland1990sampling}, $\theta_i$ can be viewed from two perspectives,
known as the ``stochastic subject" and the ``random sampling" regimes, which lead to different parameter spaces. The ``stochastic subject" regime was first adopted in the pioneering work of \cite{rasch1960probabilistic}. It treats each $\theta_i$ as an unknown model parameter to be estimated from data rather than a random sample from a certain population. This regime leads to the joint likelihood function \citep{lord1968analysis}
\begin{equation}\label{eq:jml}
L_{JL} (\boldsymbol\pi_1, ..., \boldsymbol\pi_J, \theta_1, ..., \theta_N) = \prod_{i = 1}^N \prod_{j=1}^J g_j(\theta_i \vert \boldsymbol\pi_j)^{y_{ij}}  (1-g_j(\theta_i \vert \boldsymbol\pi_j))^{1-y_{ij}}.
\end{equation}
The ``random sampling" regime assumes that $\theta_i$s are independent and identically distributed samples from a population with a density function $f$ defined with respect to a certain dominating measure $\mu$.
As a result,
one estimates the distribution function $f$ from data, instead of the individual $\theta_i$s.
This regime leads to the marginal likelihood function \citep{bock1981marginal}
\begin{equation}\label{eq:mml}
L_{ML} (\boldsymbol\pi_1, ..., \boldsymbol\pi_J, f) = \prod_{i = 1}^N \left\{\int \prod_{j=1}^J g_j(\theta \vert \boldsymbol\pi_j)^{y_{ij}}  (1-g_j(\theta \vert \boldsymbol\pi_j))^{1-y_{ij}} f(\theta) \mu d\theta\right\}.
\end{equation}
As discussed in \cite{holland1990sampling}, both regimes have their unique strengths and weaknesses but the ``random sampling" regime may be more solid as a foundation for statistical inference. The readers are referred to \cite{holland1990sampling} for a detailed discussion of the IRT assumptions from a statistical sampling viewpoint.
We will revisit these likelihood functions in Section~\ref{subsec:est_item} when the corresponding estimation problems are discussed. Specifically, we will demonstrate that, under a double asymptotic regime where both the sample size $N$ and the number of items $J$ growing to infinity, the two likelihoods converge in a certain sense. Thus, the estimates obtained from one likelihood can approximate the other. {As the ``random sampling" regime is probably more widely accepted in the literature, we will adopt this regime in the rest of the paper, unless otherwise stated.} 
We also point out that, besides these two regimes,  IRT models can be understood from a full Bayesian perspective or an ``item sampling" rationale; see \cite{thissen2009item} for further discussions.

Unidimensional IRT models discussed above can be viewed as a special case of multidimensional IRT models. In multidimensional IRT models, each individual is characterized by a vector of latent variables, denoted by $\ttt_i = (\theta_{i1}, ..., \theta_{iK})^\top$, where $K$ is the number of latent variables. The specification of a multidimensional IRT model is similar to that of a unidimensional model, except that   $g_j$ and  $f$ are multivariate. 
For example, the IRF of the multivariate two-parameter logistic (M2PL) model \citep{reckase2009multidimensional} takes the form
\begin{equation}\label{eq:m2pl}
g_j(\ttt\vert \boldsymbol\pi_j) = \frac{\exp(d_j + a_{j1}\theta_{1} + \cdots + a_{jK}\theta_{K})}{1 + \exp(d_j + a_{j1}\theta_{1} + \cdots + a_{jK}\theta_{K})},
\end{equation}
where $\ttt = (\theta_{1}, ..., \theta_K)^\top$ and $\boldsymbol\pi_j = \{a_{j1}, ...,a_{jK}, d_j\}$.
In addition, the distribution $f$ is often assumed to be multivariate normal \citep[see e.g.,][]{reckase2009multidimensional}.
Recall that we denote vectors by boldface symbols. In the rest of the paper, we will use the boldface symbol $\ttt_i$ as a generic notation for the person parameters in a general IRT model, unless the discussion is specifically about unidimensional IRT models. Moreover, with slight abuse of notation, we denote the joint and marginal likelihoods for a general IRT model as $L_{JL} (\boldsymbol\pi_1, ..., \boldsymbol\pi_J, \ttt_1, ..., \ttt_N)$ and $L_{ML} (\boldsymbol\pi_1, ..., \boldsymbol\pi_J, f)$, respectively.

As commonly seen in latent variable models  (e.g., linear factor models), suitable constraints are needed to ensure model identifiability. We take the M2PL model \eqref{eq:m2pl} as an example to illustrate this point, where $\ttt_i$ is assumed to follow a multivariate normal distribution. First, it is easy to observe that the model is not determined under
location-scale transformations of the latent variables, in the sense that one can simultaneously transform the latent variables and the corresponding loading and intercept parameters without changing the distribution of observed data.
This indeterminacy is typically resolved by letting each latent variable $\theta_{ik}$ have mean 0 and variance 1. Second, the model also has rotational indeterminacy, even after fixing the location and scale of the latent variables. That is, one can simultaneously rotate the latent variables $\ttt_i$ and the loading matrix $A = (a_{jk})_{J\times K}$ without changing the distribution of the response data. See Chapter 8, \cite{reckase2009multidimensional} for a further discussion about these indeterminacies of IRT models.

\begin{figure}
  \centering
  \includegraphics[width=0.4\textwidth]{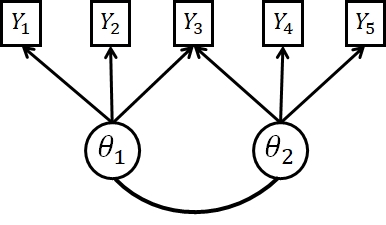}
  \caption{the path diagram of multidimensional IRT model given $Q$-matrix in equation~\ref{eq:Q}. The $Q$-matrix is indicated by the directed edges from the latent traits to the responses. The undirected edge means that the two latent traits are allowed to be correlated. The individual subscript $i$ is omitted here for simplicity.}\label{fig:Q}
\end{figure}

The rotational indeterminacy is handled differently under the confirmatory and exploratory settings of IRT analysis.
Under the confirmatory setting, design information (e.g., from the blueprint of a test)  reveals the relationship between the items and the latent variables being measured.
This design information can be coded by a $J\times K$ binary matrix, often known as the $Q$-matrix \citep{tatsuoka1983rule}, $Q = (q_{jk})_{J\times K}$, where $q_{jk} = 1$ if item $j$ directly measures the $k$th dimension and $q_{jk} = 0$ otherwise.
The $Q$-matrix imposes zero constraints on the loading parameters. That is, $a_{jk}$ will be constrained to 0 when $q_{jk} = 0$ so that $Y_{ij}$ is conditionally independent of $\theta_{ik}$ given the rest of the latent traits.
Figure~\ref{fig:Q} provides the path diagram of the multidimensional IRT model given its $Q$-matrix. As indicated by the directed edges from the latent traits to the responses,  the two latent traits are directly measured by items 1-3 and 3-5, respectively.  The $Q$-matrix thus takes the form
\begin{equation}\label{eq:Q}
Q = \left(\begin{array}{ccccc}
        1 & 1 & 1 & 0 & 0 \\
        0 & 0 & 1 & 1 & 1
      \end{array}\right)^\top.
\end{equation}
      For a carefully designed test whose $Q$-matrix satisfies suitable regularity conditions, it can be shown that the rotational indeterminacy no longer exists, as rotating the loading matrix will lead to violation of the zero constraints imposed by the $Q$-matrix \citep{anderson1956statistical,chen2019structured}.  Under the exploratory setting, the $Q$-matrix is not available. However, one would still assume, either explicitly or implicitly,  that
 the relationship between the items and the latent traits can be characterized by a sparse
$Q$-matrix and thus the corresponding loading matrix is sparse. Under this assumption, one fixes the rotation of the latent vector according to the sparsity level of the corresponding loading matrix. This problem plays an important role in the multidimensional measurement and will be further discussed in Section~\ref{subsec:structure}.

\subsection{Summary of IRT  Analysis}\label{subsec:analysis}

We describe statistical analyses under an IRT model and their purposes while leaving the more technical discussions to Section~\ref{sec:admin}. We divide the analyses into the following four categories:
\begin{enumerate}
  \item \textit{Estimation of the item-specific parameters.} The estimate can reveal the psychometric properties of each item, and thus is often used to decide whether an item should be chosen into a measurement scale,  monitor the change in items' psychometric properties over time (e.g., change due to the leakage of the item in an educational test that may be revealed by the change in the parameter estimate), among others. Estimation methods have been developed  under the ``random sampling" and ``stochastic subject" regimes; see Section~\ref{subsec:est_item}.
  \item \textit{Estimation of the person-specific parameters}. This problem is often referred to as the prediction of person parameters under the ``random sampling" regime \citep{holland1990sampling} where they are viewed as random objects. This problem is closely related to the scoring of individuals in the measurement problem. For example, in a unidimensional IRT model for educational testing where all the IRFs are monotone increasing in $\theta$, two estimates satisfying $\hat \theta_i > \hat \theta_{i'}$ implies that test-taker $i$ is predicted to answer all items more correctly than test taker $i'$. This estimate is often interpreted as an estimate of
      the test taker's ability. 
      For another example, when a multidimensional IRT model is applied to a personality test, each component of the estimate $\hat{\boldsymbol\theta_i}$ may be interpreted as an estimate of the individual's level on a certain personality trait. As will be further discussed in Section~\ref{subsec:est_person}, supposing that all the item parameters have been accurately estimated, the person parameters can be estimated based on the individual's responses to a subset of the items rather than all the test items. This is an important feature of IRT that is substantially different from the classical test theory; see a comparison between the two paradigms in Section~\ref{subsec:CTT} below. 
  \item \textit{Model evaluation.} As the psychometric interpretations of an IRT model rely on the (approximate) satisfaction of the model assumptions, these assumptions need to be checked carefully, from the evaluation of individual assumptions (e.g., the forms of IRF and marginal distribution, local independence assumption, etc.) to assessing the overall goodness-of-fit.
      Many statistical methods have been developed for evaluating IRT models; see Section~\ref{subsec:evaluation} for a discussion.
  \item \textit{Learning the latent structure of IRT models.} Like in exploratory factor analysis \citep[e.g.,][]{anderson2003introduction}, it is also often of interest to uncover the underlying structure of a relatively large number of items in multidimensional IRT models by learning the latent dimensionality and
  a sparse representation of the relationship between the latent variables and the items, i.e., the $Q$-matrix mentioned previously. The learning of the latent dimensionality is closely related to
   the determination of the number of factors in exploratory factor analysis. The $Q$-matrix learning problem  is closely related to, and can be viewed as an extension of, the analytic rotation analysis \citep{browne2001overview} in exploratory factor analysis. Besides, the
      differential item functioning problem to be discussed in Section~\ref{subsec:moreexample}  also involves learning the relationship between the observed responses, latent traits, and individuals' covariate information. Further discussions can be found in Section~\ref{subsec:structure}.
\end{enumerate}

\subsection{Comparison with Classical Test Theory}\label{subsec:CTT}

We now compare   IRT   with classical testing theory (CTT; \citealp{lord1968statistical}),
a more classical paradigm for measurement theory.
The major difference is that CTT uses the test total score to measure individuals. It
is suitable when different individuals answer the same items, but is less powerful, and sometimes infeasible,
when individuals receive different test items (e.g., in computerized adaptive testing where different test takers receive different sets of items).
More precisely, under the current notation, if all the items are equally weighted, then the total score of individual $i$ is defined as $X_i :=  \sum_{j=1}^J Y_{ij}$. The CTT decomposes the total score as
$X_i = T_i + e_i,$
where $T_i$ is the true score of person $i$ and $e_i$ is the measurement error of the current test administration satisfying $Ee_i = 0$. Conceptually, the true score $T_i$ is
defined as the expected number-correct score over an infinite number of independent administrations of the same test, hypothetically assuming that the individual does not keep the memory of test items after each administration \citep[Chapter 2,][]{lord1968statistical}. CTT further assumes that $T_i$ and $e_i$ are uncorrelated and $(T_i, e_i)$, $i = 1, ..., N$, are i.i.d. samples from a population. Under the CTT framework, the measurement of test-takers' ability becomes to estimate $T_i$s. A natural estimator of $T_i$ is the total score $X_i$.  This estimator is unbiased, since $\mathbb E(X_i \vert T_i) = T_i$.


A major contribution of CTT is to formally take the effect of measurement error into account in the modeling of testing data.
This uncertainty leads to the concept of test reliability,
defined as
$$\rho^2_{TX} := \frac{Var(T_i)}{Var(X_i)} = 1 - \frac{Var(e_i)}{Var(X_i)},$$
which reflects the relative influence of the true and error scores on attained test scores.
This coefficient is closely related to the coefficient of determination (i.e.,  R-squared) in linear regression.
However, it is worth noting that based on a single test administration and without additional assumptions, one cannot disentangle the effects of the true score and the measurement error from the observed total scores. In other words,
the true score $T_i$ is not directly observed in CTT due to its latency, unlike the dependent variables in linear regression.
Consequently, $\rho^2_{TX}$ cannot be estimated with the total scores only.
In fact, methods for the estimation of test reliability, including
Cronbach’s alpha \citep{cronbach1951coefficient}
and the split-half, test-retest, and parallel-form reliability coefficients \citep[Chapter 9,][]{lord1968statistical},  all require additional assumptions. These additional assumptions
essentially create repeated measurements of true score $T_i$.
 On the other hand, repeated measurements are automatically taken into account in an IRT model by modeling item-level data (each item as a repeated measure of the latent trait). Consequently, similar reliability coefficients are more straightforward to define and estimate under the IRT framework \citep{kim2012note}. For example, an analogous definition of $\rho^2_{TX}$ under a unidimensional IRT model is the so-called marginal reliability, defined as
$Var(E(\theta_i\vert \YY_i))/Var(\theta_i)$.


The CTT and IRT frameworks are closely related to each other. In particular, CTT can be regarded as a first-order IRT model \citep{holland2003classical}. That is, under a unidimensional IRT model given in Section~\ref{subsec:model}, the true score of a test can be defined as
$$T_i := E(X_i\vert \theta_i) = \sum_{j=1}^J g_j(\theta_i \vert \boldsymbol\pi_j).$$
Under the monotonicity assumption of the IRFs, the true score $T_i$ can be viewed as a monotone transformed latent trait level.
The model parameters $\theta_i$ and $\boldsymbol \pi_j$ can be estimated from item-level data,  leading to an estimate of the true score $T_i$.

\subsection{Connection with Linear Factor Model}\label{subsec:FA}

We next discuss the connection between IRT models and linear factor models. Specifically, we will show that these two types of models can be viewed as special cases of a general linear latent variable modeling framework \citep[GLLVM, Chapter 2,][]{bartholomew2011latent}, where IRT models focus on categorical items while linear factor models concern continuous variables. Moreover, it will be shown that by taking an underlying variable formulation \citep[Chapter 4,][]{bartholomew2011latent},  an IRT model can be viewed as a truncated version of the linear factor model.

\subsubsection{A general linear latent variable modeling framework.} The GLLVM framework was first introduced in \cite{bartholomew1984foundations}, and this framework has been
further unified and extended in a wider sense
in \cite{moustaki2000generalized} and \cite{rabe2004generalized}. Recall that an IRT model consists of (1) specification of the conditional distribution of each response $Y_{ij}$ given the latent trait, and (2) a local independence assumption and (3) an assumption on the marginal distribution of $\theta_i$.  The GLLVM specifies a general family of latent variable models following the three components (1)--(3). It allows for flexible choices of the conditional distribution of $Y_{ij}$ given $\ttt_i$ in (1) and the marginal distribution of $\ttt_i$ in (3), where the latent variables $\ttt_i$ are allowed to be unidimensional or multidimensional. In this framework, the conditional distribution of $Y_{ij}$ given $\ttt_i$ is allowed to be any generalized linear models \citep{mccullagh1989generalized}.
Specifically, one obtains a linear factor model if $Y_{ij}$ given $\ttt_i$ follows a normal distribution $N(d_j + a_{j1}\theta_{1} + \cdots + a_{jK}\theta_{K},\sigma_j^2)$ and $\ttt_i$ follows a multivariate normal distribution.
In contrast, as reviewed previously, for binary response data, $Y_{ij}$ given $\ttt_i$ follows a Bernoulli distribution with mean $g_j(\ttt_i \vert \boldsymbol\pi_j)$ in a (multidimensional) IRT model.

Most commonly used IRT models, including IRT models for other types of responses (e.g., ordinal/nominal) and multidimensional IRT models, can be viewed as special cases of the GLLVM. Such models include the partial credit model \citep{masters1982rasch}, the generalized partial credit model \citep{muraki1992generalized}, and the graded response model \citep{samejima1969estimation}, the nominal response model \citep{bock1972estimating}, and their multidimensional extensions \citep{reckase2009multidimensional}.
Under the GLLVM framework, an IRT model can be further embedded into a structural equation model, where the latent trait/traits measured by the IRT model can serve as latent dependent or explanatory variables in a structural equation model for studying the structural relationships among a set of latent and observed variables \citep{bollen1989structural}.

\subsubsection{Underlying variable formulation.}\label{subsubsec:uv} Taking an underlying variable formulation \citep{christoffersson1975factor,muthen1984general}, one can obtain a multidimensional IRT model by truncating a linear factor model. Suppose that $\tilde{\YY}_i = (\tilde Y_{i1}, ..., \tilde Y_{iJ})^\top$ follows a linear factor model with $K$ factors. Let $Y_{ij} = 1_{\{\tilde Y_{ij}\geq 0\}}$.
Then $\YY_i = (Y_{i1}, ..., Y_{iJ})^\top$, which contains truncated $\tilde Y_{ij}$, follows a $K$-dimensional IRT model. More specifically, suppose that
$\tilde Y_{ij}$ given $\ttt_i$ follows a normal distribution $N(d_j + a_{j1}\theta_{1} + \cdots + a_{jK}\theta_{K},1)$. Then $P(Y_{ij} = 1\vert \ttt_i ) = P(\tilde Y_{ij} \geq 0\vert \ttt_i) = \Phi(d_j + a_{j1}\theta_{1} + \cdots + a_{jK}\theta_{K})$. Recall that $\Phi$ is the cumulative distribution function of the standard normal distribution. Together with  the assumptions of local independence and the marginal distribution $f$, it specifies a multidimensional IRT model under the probit link function. This formulation can be easily extended for ordinal response data.

Assuming normality in both the conditional distribution of $\tilde{\YY}_i$ given $\ttt_i$ and the marginal distribution of $\ttt_i$, the underlying variable formulation is closely related to the tetrachoric/polychoric correlations for measuring the association between binary/ordinal variables, which dates back to the seminal work of \cite{pearson1900mathematical}.
As will be discussed in Section~\ref{subsec:est_item}, this connection leads to a computationally efficient method for estimating the corresponding multidimensional IRT models.

\subsection{Connection with Analysis of Contingency Tables}\label{subsec:contingency}

IRT models have a close relation with categorical data analysis,
particularly the analysis of large sparse multidimensional contingency tables; see  \cite{fienberg2000contingency}.
In fact, response data for $J$ items can be regarded as a $J$-way contingency table with
$2^J$ cells, where each cell records the total count of a response pattern (i.e., a binary response vector). Note that the
$J$-way contingency table is a sufficient statistic for the raw response data.
This contingency table is typically sparse, as the sample size $N$ is usually much smaller than the number of cells when the number of items $J$ is moderately large (e.g., $J \geq 20$). Even when the sample size and the number of cells are comparable, some cells can still be sparse since the counts in the cells may be highly dependent on each other. Parsimonious models have been proposed for analyzing  high-way contingency tables, which impose sparsity in the coefficients for
higher-order interaction terms.

As pointed out in \cite{holland1990dutch}, IRT models approximate  the second-order loglinear model for $J$-way contingency tables
\begin{equation}\label{eq:soe}
P(\YY_i = \yy) \propto \exp\left(\mathbf d^\top \yy + \frac{1}{2}\yy^\top AA^\top \yy\right),
\end{equation}
where $\mathbf d = (d_1, ..., d_J)^\top$ and $A = (a_{jk})_{J\times K}$ ($K$ is often chosen to be much smaller than $J$, leading to a parsimonious model). This model
 does not explicitly contain latent variables.  It is also known as an
Ising model \citep{ising1925beitrag},
an important model in statistical mechanics. Note that the same notations are used in the second-order loglinear model as those in the M2PL model in \eqref{eq:m2pl}, for reasons explained below. That is, the second-order loglinear model can be viewed as an M2PL model with a special marginal distribution for the latent variables. More precisely, if the joint distribution of $(\YY_i, \ttt_i)$ takes the form
\begin{equation}\label{eq:soe2}
f(\yy, \ttt) \propto \exp\left(-\frac{1}{2}\ttt^\top\ttt + \yy^\top A\ttt +  \mathbf d^\top \yy \right),
\end{equation}
then integrating out $\ttt$ gives the second-order log-linear model \eqref{eq:soe}.
Furthermore, it can be easily shown that the conditional distribution of $\YY_i$ given $\ttt_i$  is the same as that of the M2PL model given by equation
\eqref{eq:m2pl}. Under the joint model \eqref{eq:soe2}, the marginal distribution of $\ttt_i$ becomes a Gaussian mixture.
For more discussions on the connection between IRT models and loglinear models, we refer readers to \cite{fienberg1983loglinear}, \cite{ip2002locally},  and \cite{tjur1982connection}.

\subsection{Comparison with Generalized Linear Mixed Models}\label{subsec:glmm}

The specification of IRT models under the ``random sampling" regime
is similar to that of generalized linear mixed models  \citep[GLMM;][]{berridge2011multivariate,searle2001generalized}, and many IRT models can be viewed as special cases of GLMM. GLMM extends the generalized linear model to analyzing grouped (i.e., clustered) data commonly seen in longitudinal or repeated measures designs.
A GLMM adds a random effect into a generalized linear model (e.g., logistic regression model) while keeping the fixed effect in the generalized linear model for studying the relationship between observed covariates and an outcome variable.
The random effect is used to  model the dependence among outcome variables within the same group (cluster).
The IRT models introduced in Section~\ref{subsec:model} can all be viewed as special cases under the GLMM framework. Specifically,
each individual can be viewed as a group, and the individual's item responses can be viewed as repeated measures. The latent variables $\ttt_i$  correspond to the random effects in the GLMM. There are no observed covariates in these basic models but  there are more complex IRT models to be discussed in Section~\ref{subsec:moreexample} below that make use of individual-specific and item-specific covariates. Due to the similarities between IRT models and GLMMs, the estimation methods to be discussed in Section~\ref{subsec:est_item} for IRT models also apply to GLMMs.

While the two families of models largely overlap with each other in terms of the model assumptions, their main purposes are different, at least in the historical applications of these models. That is, the GLMM focuses on explaining data by
testing certain hypotheses about the fixed effect parameters (i.e., the regression coefficients for observed covariates), treating the random effect as
a component of the model that is not of interest in the statistical inference but necessary for capturing the  within group-dependence.
On the other hand, IRT models tend to focus on measuring individuals. Thus, the inference about the latent variables $\ttt_i$, the random effect from the GLMM perspective,  is
of particular interest. However, it is worth pointing out an important family of IRT models, called the explanatory IRT models \citep{de2004explanatory}, that combines the explanatory perspective of GLMM and the measurement perspective of IRT. These models are specified under the GLMM framework, incorporating person- and item-specific covariates to explain the characteristics of individuals and items, respectively.

\subsection{Connection with Collaborative Filtering and Matrix Completion}
IRT models are also closely related to collaborative filtering  \citep{koren2015advances,zhu2016personalized},   a method of making automatic predictions (filtering) about the interests of a user by collecting preference/taste information from many users (collaborating) that is widely used in recommendation systems (e.g., e-commerce).
The user-by-item matrix in collaborative filtering can be viewed as an item response data matrix,
for which a large proportion of entries are missing due to the nature of the problem.
A famous example of collaborative filtering is the Netflix challenge on movie recommendation \citep{feuerverger2012statistical}.  Data of this challenge are ratings from a large number of users to a large set of movies, where many missing values exist as each user only watched a relatively small number of movies. The goal is to learn the preferences of each user on movies that they have not watched. The collaborative filtering problem, including the Netflix example,
can be cast into a matrix completion problem that concerns filling the missing entries of a partially observed matrix.
When the entries of the data matrix are binary or categorical-valued, the problem is known as a one-bit or categorical matrix completion problem.

Without further assumptions, the matrix completion problem is ill-posed since the missing entries can be assigned arbitrary values. To create a well-posed problem, a low-rank assumption is typically imposed to reduce the number of parameters, which is similar to the introduction of low-dimensional latent variables in IRT models. More precisely, for the completion of a binary or categorical matrix, it is often assumed that the data matrix follows  a probabilistic model parameterized by a low-rank matrix \citep{bhaskar2016probabilistic,bhaskar20151,cai2013max,davenport20141,zhu2016personalized}. From the statistical sampling perspective, these models
take the ``stochastic subject" regime that treats the user-specific parameters as fixed parameters.
The specifications of such models are very similar to multidimensional IRT models.
In particular, the one-bit matrix completion problem aims to recover the matrix $(p_{ij})_{N\times J}$, where $p_{ij} = P(Y_{ij} = 1 \vert \ttt_i ) = g_j({\ttt}_i \vert {\boldsymbol\pi}_j)$. A major difference between collaborative filtering and psychometric applications of IRT is that psychometric applications are typically interested in the inference of the latent variables and the item parameters, while collaborative filtering only focuses on the inference of $p_{ij}$s.
As will be discussed in Section~\ref{subsec:prediction}, some psychometric problems may be viewed as collaborative filtering problems, and may be solved efficiently by matrix completion and related algorithms.

\subsection{More Complex IRT Models and Their Psychometric Applications}\label{subsec:moreexample}

In what follows, we review several more complex IRT models beyond the basic forms given in Section~\ref{subsec:model}, and discuss their psychometric applications. Note that all these models can be viewed as special cases under the GLLVM framework discussed in Section~\ref{subsec:FA}.

\subsubsection{IRT models involving covariates.} Sometimes, covariates of the individuals are collected together with item response data. We use $\xx_i$ to denote $p$-dimensional observed covariates of individual $i$. Covariate information can be incorporated into the IRT model in different ways for different purposes. We discuss two types of models often used in psychometrics.

\begin{figure}
  \centering
  \includegraphics[width= 0.8\textwidth]{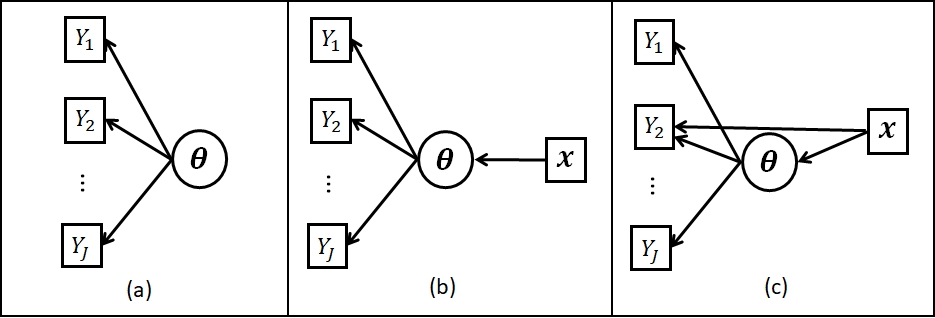}
  \caption{Path diagrams for three IRT models, including (a) an IRT model without covariates, (b) a latent regression model, and (c) an MIMIC model. The individual subscript $i$ is omitted here for simplicity.}\label{fig:models}
\end{figure}

First, the covariates may  affect the distribution of the latent traits, but not directly on those of the item responses. Such models are known as the latent regression IRT models \citep{mislevy1984estimating,mislevy1985estimation}. They are useful in large-scale assessments, such as PISA and TIMSS, for estimating group-level distributions of the corresponding latent traits for policy-relevant sub-populations such as gender and ethnicity groups.  For example, covariate information can be incorporated into a unidimensional IRT model (e.g., 2PL model) by assuming  $\theta_i$ to follow a normal distribution $N(\xx_i^\top \boldsymbol\beta, 1)$ instead of a standard normal distribution, where $\boldsymbol\beta$ is a $p$-dimensional vector of the coefficients. Note that
the rest of the model assumptions remain the same (i.e., the IRFs and the local independence assumption). In particular, the covariates are not involved in the IRFs, so that the covariates do not directly affect the distribution of item responses. This model implies that the mean of the latent trait distribution depends on the covariates. When the covariates are indicators of group memberships (e.g., gender, ethnicity), the model allows the group means to be different. Figure~\ref{fig:models} provides a graphical representation of the latent regression IRT model and other relevant models. Specifically, panels (a) and (b) of the figure show the path diagrams for a basic IRT model and a latent regression IRT model, respectively. In panel (b), the arrow from $\xx$ to $\ttt$ implies that the distribution of $\ttt_i$ depends on the covariates and the absence of an arrow from
$\xx$ to the responses  implies that the covariates do not directly affect the distribution of the response variables given the latent traits. The latent regression models can also be regarded as a special explanatory IRT model  \citep{de2004explanatory}; see Section~\ref{subsec:glmm} for a brief discussion of explanatory IRT models.

Second,  the covariates may affect  the distributions of the  latent traits and the responses to some items. Such models are typically known as the Multiple Indicators, Multiple Causes (MIMIC) models, originally developed in the literature of structural equation models \citep{goldberger1972structural,robinson1974identification,zellner1970estimation} and then extended to IRT models \citep{muthen1985method,muthen1988some}
for studying differential item functioning (DIF). DIF refers to the situation when items may function differently for different groups of individuals (e.g., gender/ethnicity)
or they may measure even different things for members of one group as opposed to members of another. For example, a reading comprehension item in a language test may show DIF between the male and female groups if the content of the reading paragraphs is about a specific topic that either males or females have more experience with. The MIMIC model provides a nice way to describe the DIF phenomenon. For simplicity, consider the case of a single binary covariate $x_i \in {\{0, 1\}}$ that indicates the group membership. Then MIMIC model let the IRFs of the DIF items depend on the group membership, while allowing the two groups to have different latent trait distributions. For example, suppose that item $j$ is the only DIF item among $J$ items. Then a simple MIMIC model for DIF can be set as follows.
The IRF of item $j$ can be modeled as
\begin{equation}\label{eq:2plDIF}
\frac{\exp(d_j + a_j \theta + \gamma_j x_i)}{1+\exp(d_j + a_j \theta + \gamma_jx_i)},
\end{equation}
which takes a 2PL model framework with $\gamma_j$ being the parameter characterizing the group effect on the IRF.
The rest of the items are DIF-free items, and thus, their
IRFs still take the standard 2PL form \eqref{eq:2pl}.
The latent trait distribution can be modelled by setting $\theta_i$ given $x_i $ to follow $N(\beta x_i, 1)$. Panel (c) of Figure~\ref{fig:models} gives the path diagram for a MIMIC model, where the second item is a DIF item.
In real-world DIF analysis, DIF items are unknown and need to be detected based on data. As will be discussed in Section~\ref{subsec:structure}, the DIF analysis can be cast into a model selection problem. Comparing with other statistical model selection problems, such as variable selection in regression models, the current model has
additional location-shift indeterminacy brought by the introduction of covariates, which complicates the analysis.






\subsubsection{Discrete latent variables.} The latent variables in an IRT model can also be discrete, though in many commonly used models, especially unidimensional models, they are assumed to be continuous variables.  IRT models with categorical latent variables are known as latent class models \citep{goodman1974analysis,goodman1974exploratory,haberman1977product,lazarsfeld1950logical,lazarsfeld1959latent,lazarsfeld1968latent}. In the context of psychological and educational measurement, a latent class model is preferred when the goal of measurement is to classify people into groups, while an IRT model with continuous latent variables is preferred if the goal is to assign people scores on a continuum.
In particular, the diagnostic classification models (DCMs),
also known as the cognitive diagnosis models, are a family of restricted latent class models dating back to the works of \cite{haertel1989using}  and \cite{macready1977use}. DCMs have received much recent attention in educational assessment, psychiatric evaluation, and many
other disciplines \citep{templin2010diagnostic,vondavier2019handbook}. In particular, such models have played an important role in intelligent tutoring systems \citep{cen2006learning,doignon2012knowledge} to support personalized learning.
In DCMs, an individual is represented by a latent vector $\ttt_i = (\theta_{i1}, ..., \theta_{iK})^\top$,
each component of which is a
discrete latent variable that represents the individual's status on a certain attribute. For example, in the binary case, $\theta_{ik} = 1$ may represent the mastery of skill $k$, and $\theta_{ik} = 0$ otherwise. 
As a result, a diagnostic classification model classifies
each individual given their item responses into a latent class that represents the individual's profile on multiple discrete attributes.

Measurement based on DCMs is multidimensional and confirmatory in nature, as these models are used to
measure multiple fine-grained constructs of individuals simultaneously. As with the multidimensional IRT models discussed in Section~\ref{subsec:model}, the $Q$-matrix also plays an important role in the DCMs. Recall that
the $Q$-matrix takes the form $Q = (q_{jk})_{J\times K}$, where each entry $q_{jk}$ is binary, indicating
whether item $j$ directly measures the $k$th dimension or not.
Given the $Q$-matrix, different
IRFs have been developed to capture the way a set of relevant attributes affect an item. For example, the Deterministic Inputs,
 Noisy ``And" gate (DINA) model \citep{junker2001cognitive} is a popular diagnostic classification model in educational measurement. The IRF of this model assumes that
an individual will answer an item correctly provided that the individual acquires all the relevant skills,
subject to noise due to
slipping (when the individual can answer correctly) and guessing (when the individual is not able to answer correctly) behaviors.
More precisely, the DINA model assumes binary latent variables (i.e., $\theta_k \in \{0, 1\}$) and the following IRF,
\begin{equation}\label{eq:dina}
  g_j(\ttt\vert \boldsymbol\pi_j) = \left\{\begin{array}{ll}
                         1 - s_j &\mbox{~if~} \theta_k \geq q_{jk}, \mbox{~for all~} k = 1, ..., K,\\
                         g_j  & \mbox{~otherwise}, \\
                       \end{array}\right.
\end{equation}
where $s_j$ and $g_j$ are two item-specific parameters capturing the chances of answering incorrectly due to carelessness when he/she is able to solve the problem and of guessing the answer correctly, respectively, and $\boldsymbol\pi_j = (s_j, g_j)$. The indicating event in \eqref{eq:dina}, $1_{\{\theta_k \geq q_{jk},  k = 1, \cdots, K\}}$ is also known as the ideal response. 
Several DCMs have been developed to account for different types of psychological processes underlying item response behavior, such as the Deterministic Inputs,
 Noisy ``Or" gate (DINO) model \citep{templin2006measurement} and reparametrized unified models \citep{dibello1995unified,junker2001cognitive}, for which different forms of IRFs are assumed. All these DCMs can be viewed as special cases of
a general diagnostic classification model \citep{de2011generalized,henson2009defining,von2008general}, where the IRF takes the general form containing all the interactions of latent variables
\begin{equation}\label{eq:dcms}
\begin{aligned}
h_j^{-1}(g_j(\ttt\vert \boldsymbol\pi_j)) =& a_{j0} + a_{j1}q_{j1}\theta_1 + \cdots + a_{jK}q_{jK}\theta_K \\
+& a_{j12}q_{j1}q_{j2} \theta_1 \theta_2 + \cdots + a_{j,K-1,K-2} q_{j,K-1}q_{j,K} \theta_{K-1} \theta_K \\
+& \cdots +  a_{j,1,\cdots,K} \prod_{k=1}^K q_{jk}\theta_{k}.
\end{aligned}
\end{equation}
In \eqref{eq:dcms}, $h_j$ is a link function (e.g., $h_j(x) = \exp(x)/(1+\exp(x))$), and the item parameter vector $\boldsymbol \pi_j = \{a_{j0}, a_{j1}, \cdots,  a_{j,1,\cdots,K}\}$.
Note that the latent variable interactions are important to capture the
disjunctive or conjunctive relationships between the latent attributes in the special cases such as the DINA and DINO models; see, e.g., \cite{de2011generalized} for more details.

Fundamental identifiability issues arise with the relatively more complex latent structure in DCMs. The interpretation and measurement of the latent constructs are only valid under an identifiable model. The identifiability problem of DCMs, and more generally of multidimensional IRT models, has two levels. First, under what measurement design are the model parameters ($\boldsymbol\pi_j$ and parameters in $f$, the marginal distribution of $\ttt$) identifiable, assuming that the $Q$-matrix is known and correctly specified? Second, if the $Q$-matrix is unknown, when can we simultaneously identify the model parameters and the $Q$-matrix from data? Efforts have been made to address these questions; see \cite{chen2015statistical},  \cite{chen2019structured},
\cite{chiu2009cluster}, \cite{fang2019identifiability}, \cite{fang2020identifiability}, \cite{gu2020partial}, \cite{liu2013theory},  \cite{xu2016identifiability}, \cite{xu2017identifiability}, and \cite{xu2018identifying}.

%
%

\subsubsection{Nonparametric IRT models.} Nonparametric modeling techniques have been incorporated into
IRT, yielding more flexible models.
They play an important role in assessing the goodness-of-fit of parametric IRT models and providing robust measurement against model misspecification.

Under the ``random sampling" regime, one can assume either the IRF $g_j(\ttt\vert \boldsymbol\pi_j)$ or the distribution $f$ to be nonparametric.
However, we note that model non-identifiability generally occurs if
assuming both to be nonparametric. In that case, one can simultaneously transform the IRFs and the latent variable distribution without changing the distribution of item responses; see \cite{ramsay1991maximum} for a discussion.
Even with only one of the IRFs and the marginal distribution $f$ being nonparametric, model
identifiability  can still be an issue. This is because, for binary response data, the sufficient statistic (i.e., count of each of $2^J-1$ response patterns) is of dimension $2^{J}-1$, which does not grow for a fixed $J$. At the same time, there are infinitely many parameters in the nonparametric model component. See
\cite{douglas2001asymptotic} for a discussion about the identifiability of nonparametric IRT models and theory for the asymptotic identifiability when both the sample size $N$ and the number of items $J$ grow to infinity.

 \cite{cressie1983characterizing} studied the identifiability of a semiparametric Rasch model, where the IRFs follow the form of a Rasch model and the marginal distribution $f$ is nonparametric. For the same model, \cite{de1986maximum} and \cite{lindsay1991semiparametric} further discussed the model identifiability and proposed a nonparametric marginal maximum likelihood estimator in the sense of \cite{kiefer1956consistency}, assuming $f$ to be a mixture distribution with unspecified weights at unknown points. Their results suggest that, under suitable conditions, the estimation of the item parameters is consistent under the asymptotic regime where $J$ is fixed and $N$ goes to infinity, even though there are more parameters than the sample size. \cite{haberman2005identifiability} extended the analysis by considering IRFs to take the 2PL and 3PL forms and found that the good properties of the nonparametric marginal maximum likelihood estimator for the semiparametric Rasch model do not carry over due to model non-identifiability.
There have been other developments in the nonparametric modeling of item response functions.
In these developments,  the item response functions $g_j(\ttt\vert \boldsymbol\pi_j)$ are replaced
by nonparametric functions. 
Generally speaking, research in this direction makes minimum assumptions on the item response functions, except for certain monotonicity or smoothness assumptions.
Nonparametric function estimation methods have been applied to the estimation of nonparametric IRFs,
such as spline methods; see \cite{johnson2007modeling},
\cite{ramsay1991maximum}, and \cite{winsberg1984fitting}.


Many developments on non-parametric IRT have been made under the ``stochastic subject" regime, either explicitly or implicitly, where the person parameters are treated as fixed parameters. Thus, only the IRFs are considered non-parametric. Specifically, Mokken scale analysis \citep{mokken1971theory,mokkan1982nonparametric,sijtsma2002introduction}, a pioneer work on non-parametric IRT, is developed under this regime. More precisely, a Mokken scale analysis model is a unidimensional IRT model,  assuming that each IRF is a non-parametric monotonically nondecreasing function of the latent trait. Sometimes, it further makes the ``non-intersecting IRFs" assumption  that imposes a monotone ordering of IRFs. See \cite{sijtsma2017tutorial} for a discussion of these monotonicity assumptions. More general non-parametric IRT models have been proposed, for which theory and estimation methods have been developed. For an incomplete list
of these developments, see \cite{douglas1997joint}, \cite{guo2011nonparametric}, \cite{johnson2006nonparametric}, \cite{ramsay1989binomial},  \cite{sijtsma2002introduction},  and \cite{strout1990new}.
We point out that these ``stochastic subject" models have similar identifiability issues as those under the ``random sampling" regime.
Therefore, restrictions on the model are needed to resolve the indeterminacies, and a double asymptotic regime is typically needed, i.e., both $N$ and $J$ growing to infinity, for establishing consistent estimation of the non-parametric IRFs \citep[e.g.][]{douglas1997joint}.


Finally, we point out that essentially all the existing works on non-parametric IRT models focus  on unidimensional models. Non-parametric multidimensional IRT models remain to be developed, given that multidimensional measurement problems become increasingly more common these days. It is worth noting that
deep autoencoder models, a family of deep neural network models, are essentially nonlinear and non-parametric factor models \citep[Chapter 14,][]{goodfellow2016deep}. These models are flexible and control the latent dimension  through the architecture of the hidden layers in the deep neural network.
In particular, the M2PL model given in Section~\ref{subsec:model} can be viewed as a special autoencoder model with one hidden layer. With this close connection, models and algorithms for  deep autoencoders may be borrowed to develop non-parametric multidimensional measurement models.

\section{Statistical Analysis under IRT Framework}\label{sec:admin}

In this section, we discuss the statistical analyses under the IRT framework as listed in Section~\ref{subsec:analysis}.

\subsection{Estimation of Item Parameters}\label{subsec:est_item}

We first consider the estimation of item parameters, which is often known as item calibration. This estimation problem is closely related to the estimation of fixed parameters in GLMMs and other nonlinear factor models,
for which
 the methods reviewed below are also generally suitable.  We discuss these estimation methods
under the ``random sampling" and ``stochastic subject" regimes. As will be further discussed in Section \ref{subsubsec:jml} below, the two regimes converge in a certain sense when both the sample size and the number of items grow to infinity. In this double asymptotic sense, estimation under the ``stochastic subject" regime can be viewed as an approximation to that under the ``random sampling" regime,
if one believes that the latter is more solid  philologically as a foundation for statistical inference.

\subsubsection{Marginal maximum likelihood estimation.}

The ``random sampling" regime treats the person parameters as samples from a distribution. The marginal maximum likelihood (MML) estimator is the main estimation method under this regime and is also the most commonly used method in  modern applications of item calibration.
Suppose that the indeterminacies of an IRT model have been removed by imposing constraints on model parameters so that the model is identifiable.
The MML simultaneously estimates the item parameters and the distribution of person parameters by maximizing the marginal likelihood function. That is,
\begin{equation}\label{eq:mmlopt}
(\hat{\boldsymbol\pi}_1, ..., \hat{\boldsymbol\pi}_J, \hat f) =\argmax_{\boldsymbol\pi_1, ...,\boldsymbol\pi_J, f} \log L_{ML} (\boldsymbol\pi_1, ..., \boldsymbol\pi_J, f).
\end{equation}
Here, we assume the distribution $f$ to take a parametric form.
This estimator can be viewed as an empirical Bayes estimator \citep{efron1973stein,robbins1956empirical}; see \cite{efron2003robbins} and \cite{zhang2003compound} for a review of empirical Bayes methods.

{It is worth emphasizing that linking is automatically performed in the MML estimator \eqref{eq:mmlopt}, as well as some of the estimators reviewed later including the JML estimator. Suppose that each individual $i$ is only given a subset of the test items. This design is commonly used in large-scale assessments such as PISA and TIMSS to achieve good content coverage without requiring each individual to answer too many items. That is, let $B_i \subset \{1, ..., J\}$ be the set of items assigned to individual $i$. Then the marginal likelihood function can be written as
\begin{equation}\label{eq:mml2}
L_{ML} (\boldsymbol\pi_1, ..., \boldsymbol\pi_J, f) = \prod_{i = 1}^N \left\{\int \prod_{j\in B_i} g_j(\ttt \vert \boldsymbol\pi_j)^{y_{ij}}  (1-g_j(\ttt \vert \boldsymbol\pi_j))^{1-y_{ij}} f(\ttt)\mu d\ttt\right\},
\end{equation}
When there is sufficient overlap between the subsets $B_i$ and given the standard identifiability constraints for the IRT model,   the item parameters and marginal distribution $f$ can still be identified based on the marginal likelihood \eqref{eq:mml2}. Consequently, the item parameter estimates from the MML estimator based on \eqref{eq:mml2} automatically lie on the same scale. We refer the readers to \cite{mislevy1996missing} for further discussions on the application of IRT models to item response data with missing values.
}

The optimization problem~\eqref{eq:mmlopt} is often solved by an Expectation-Maximization (EM) algorithm \citep{bock1981marginal,dempster1977maximum}. When a multidimensional IRT model is fitted, the EM algorithm for solving \eqref{eq:mmlopt} is usually slow. This is because the algorithm has to frequently solve $K$-dimensional numerical integrals, whose complexity increases  exponentially with $K$. To speed up the EM algorithm, stochastic versions of the EM algorithm have been proposed.  These algorithms avoid the numerical integration by Monte Carlo simulation \citep{cai2010high,cai2010metropolis,diebolt1996stochastic,
ip2002single,meng1996fitting,zhang2018improved,zhang2020computation}.
Among these developments, we draw  attention to the stochastic approximation methods, also known as the stochastic gradient descent methods, proposed in
\cite{cai2010high,cai2010metropolis} and \cite{zhang2020computation}, which date back to the seminal work of \cite{robbins1951stochastic} on stochastic approximation and the work of \cite{gu1998stochastic} that combines
Markov chain Monte Carlo (MCMC) sampling and stochastic approximation for estimating latent variable models.

We now demonstrate how the optimization problem~\eqref{eq:mmlopt} can be solved by stochastic approximation. To simplify the notation, we use $\Xi$ to denote all the fixed parameters in $\boldsymbol\pi_1, ..., \boldsymbol\pi_J$ and  $f$, and write the log marginal likelihood function as $l(\Xi)$.
We further denote
$$l_i(\ttt_i, \Xi) =  \log \left(\prod_{j=1}^J g_j(\ttt_i \vert \boldsymbol\pi_j)^{y_{ij}}  (1-g_j(\ttt_i \vert \boldsymbol\pi_j))^{1-y_{ij}} f(\ttt_i)\right)$$
as the log complete-data likelihood for individual $i$ that is based on the joint distribution of $\ttt_i$ and $\YY_i$.
Then it can be shown that the gradient of $l(\Xi)$ takes the form
$$\nabla l(\Xi) = \sum_{i=1}^N E\left(\frac{\partial l_i(\ttt_i, \Xi)}{\partial\Xi}\Big\vert \YY_i=\yy_i\right),$$
where the conditional expectation is with respect to $\ttt_i$ given $\YY_i$. With this observation, the stochastic approximation methods \citep{cai2010high,cai2010metropolis,zhang2020computation} for solving \eqref{eq:mmlopt} iterate between two steps: (1) sample $\ttt_i$ from its conditional distribution given $\YY_i$, where the conditional distribution is based on $\Xi^{(t)}$, the current value of $\Xi$, and (2) given the obtained samples $\ttt_i^{(t)}$, $i = 1, ..., N$, update the value of $\Xi$ by a stochastic gradient ascent, where the stochastic gradient at $\Xi^{(t)}$ is given by
$$ \sum_{i=1}^N \frac{\partial l_i(\ttt_i^{(t)}, \Xi)}{\partial\Xi}\vert_{\Xi = \Xi^{(t)}}$$
whose conditional expectation (given observed responses) is $\nabla l(\Xi)$.
For multidimensional IRT models with many latent traits, it may not be straightforward to sample $\ttt_i$ from the conditional distribution, and MCMC methods are needed to perform the sampling step. Under mild conditions,
$\Xi^{(t)}$ is guaranteed to converge to the solution of optimization problem \eqref{eq:mmlopt}, even when the samples in the sampling step are approximated by an MCMC algorithm \citep{zhang2020computation}. Typical of stochastic approximation, the performance of these algorithms is sensitive to the step size in the stochastic gradient ascent step, where the step size is required to decay to 0 at a suitable rate to ensure convergence.
\cite{cai2010high,cai2010metropolis} suggested to set the step size to delay at the rate $1/t$, which is known to be asymptotically optimal for the Robbins-Monro algorithm \citep{chung1954stochastic,lai1979adaptive}.
However, the $1/t$ rate is well-known to yield unstable results in practice as it  decays to zero too fast.
\cite{zhang2020computation} suggested to use a slower-decaying step size and the Polyak–Ruppert averaging procedure \citep{polyak1992acceleration,ruppert1988efficient} to improve the empirical performance of the stochastic approximation algorithm while maintaining a fast theoretical convergence rate.




\subsubsection{Limited-information estimation methods.}\label{subsubsec:lie}

Alternative estimation methods are  developed under the ``random sampling" regime to bypass
the  high-dimensional integrals in the MML. These methods are known as limited-information methods. They  do not consider the complete joint contingency table of
all items, but only marginal tables up to a lower order (e.g., two-way tables based on item pairs).

We divide these methods into two categories. The first concerns the probit models discussed in Section~\ref{subsubsec:uv} where the IRFs take a probit form, and marginal distribution of the latent traits is also a multivariate normal distribution. Making use of the underlying variable formulation, it can be shown that these IRT models can be estimated by first estimating the multivariate normal distribution of the underlying variables and then recover the IRT parameters based on the estimated distribution of the underlying variables.
Note that the first step can be done efficiently using the one-way and two-way tables based on all the individual items and  item pairs \citep{muthen1984general}, and the second step solves an optimization problem with no integrals involved. Consequently, this method can computationally efficiently estimate IRT models with many latent traits, especially when the number of items is not too large. Developments in this direction   include \cite{christoffersson1975factor}, \cite{lee1990three,lee1992structural,lee1995two}, and \cite{muthen1978contributions,muthen1984general}. 

The second category of methods makes use of the composite likelihood method \citep{besag1974spatial,cox2004note,lindsay1988composite}. These methods
allow for more general forms of the IRFs but still require the marginal distribution of the latent traits to be normal. More specifically, the fixed parameters are estimated by maximizing a composite likelihood based on the lower-order marginal tables. In particular, the pairwise likelihood is most commonly used that is constructed based on item pairs, though more general composite likelihood functions can be constructed based on triplets or quadruplets of items; see
\cite{joreskog2001factor},  \cite{katsikatsou2012pairwise}, \cite{vasdekis2012composite}, and \cite{vasdekis2014weighted}.

Although these limited-information methods are computationally less demanding than the MML approach, there are some limitations. First, as mentioned previously, these methods only apply to some restricted classes of IRT models (e.g., probit IRFs, multivariate normal distribution for the latent traits). They thus are not as generally appliable as the MML approach. Second, when the IRT model is correctly specified, the limited-information methods
suffer from some information loss and thus
 are statistically less efficient than the MML estimator.



%

\subsubsection{Joint maximum likelihood estimation.}\label{subsubsec:jml}

The joint maximum likelihood (JML) estimator refers to the estimation method that simultaneously estimates the item and person parameters by maximizing the joint likelihood function introduced in Section~\ref{subsec:model}.
This approach was first suggested in \cite{birnbaum1968some}, and has been used in item
response analysis for many years \citep{lord1980applications,mislevy1989consumer,wood1978logist}
until the MML approach becomes dominant.
Since the joint likelihood function does not involve integrals, the computation of the JML estimator is typically much faster than that of the corresponding MML estimator, especially when the latent dimension is high.
Despite the computational advantage, the JML estimator is still less preferred to the MML for item calibration, possibly due to two reasons.
The first one is philosophical. As pointed earlier, the JML estimator naturally fits the ``stochastic subject" regime, which, however, is less well accepted than the ``random sampling" regime.  The second reason is that the JML estimator
lacks desirable asymptotic properties. Under the standard asymptotic regime where the number of item is fixed and the number of people goes to infinity, the JML estimation of the item parameters is inconsistent. This inconsistency is due to that the sample size and the dimension of parameter space  grow at the same speed. This phenomenon was originally noted in
\cite{neyman1948consistent} under a linear model and further discussed by \cite{andersen1970asymptotic} and \cite{ghosh1995inconsistent} under IRT models. 

While these reasons are valid under a setting when $J$ is small, they may no longer be a concern under a large-scale setting when both the sample size $N$ and the number of items $J$ are large. We provide some justifications for joint-likelihood-based estimation, using the double asymptotic regime that both  $N$ and   $J$ grow to infinity. We first point out that the two likelihood functions tend to approximate each other under this regime. To see this, we do local expansion of the marginal likelihood function at the MML estimator $(\hat{\boldsymbol\pi}_1, ..., \hat{\boldsymbol\pi}_J, \hat f)$. Under the double asymptotic regime and by
making use of the Laplace approximation for integrals, $\log L_{ML} (\hat {\boldsymbol\pi}_1, ..., \hat {\boldsymbol\pi}_J, \hat f)$ can be approximated by
$\log L_{JL} (\hat {\boldsymbol\pi}_1, ..., \hat {\boldsymbol\pi}_J,  \hat \ttt_1, ..., \hat \ttt_N)$
plus some smaller-order terms, where $\hat {\boldsymbol\pi}_j$ is the same MML estimator and
 $\hat{\ttt}_i$ is given by
\begin{equation*}
\begin{aligned}
\hat{\ttt}_i 
 = \argmax_{\ttt} ~\log \left(\prod_{j=1}^J g_j(\ttt \vert \hat{\boldsymbol\pi}_j)^{y_{ij}}  (1-g_j(\ttt \vert \hat{\boldsymbol\pi}_j))^{1-y_{ij}} \right).\\
\end{aligned}
\end{equation*}
See \cite{huber2004estimation} for more details about this expansion.

We further point out that some notion of consistency can be established for the JML approach under the double asymptotic regime. Specifically,
\cite{haberman1977maximum} showed that
person and item parameters of a Rasch model can be consistently estimated when $N$ and $J$ grow to infinity at a suitable rate.
\cite{chen2021note} extended the analysis of  \cite{haberman1977maximum} under a setting that many entries of the response data matrix are missing.
\cite{chen2018joint,chen2019structured} considered a suitably constrained JML estimator for more general multidimensional IRT models and showed that
the estimator achieves the optimal rate under this asymptotic regime. Specifically, the constrained JML estimator solves a JML problem with constraints on the magnitudes of person and item parameters
\begin{equation}\label{eq:cjmle}
\begin{aligned}
(\hat \ttt_1, ..., \hat \ttt_N,\hat{\boldsymbol\pi}_1, ..., \hat{\boldsymbol\pi}_J) =&\argmax_{\ttt_1, \cdots \ttt_N, \boldsymbol\pi_1, ...,\boldsymbol\pi_J} L_{JL} (\boldsymbol\pi_1, ..., \boldsymbol\pi_J, \ttt_1, ..., \ttt_N)\\
s.t. ~&\Vert\ttt_i\Vert \leq C, \Vert \boldsymbol\pi_j \Vert \leq C, i = 1, ..., N, j = 1, ..., J,
\end{aligned}
\end{equation}
where $C$ is a pre-specified constant. Under mild conditions and assuming a fixed latent dimension $K$, \cite{chen2019structured} showed that
\begin{equation}\label{eq:optrate}
\frac{\Vert \hat P - P^* \Vert_F}{\sqrt{NJ}} = O_p\left(\frac{1}{\sqrt{N\wedge J}}\right)
\end{equation}
is the optimal rate (in minimax sense) for estimating $P^*$.
Here, $\hat P = (g_j(\hat \ttt_i \vert \hat{\boldsymbol\pi}_j))_{N\times J}$ and $P^* = (g_j(\ttt_i^* \vert {\boldsymbol\pi}_j^*))_{N\times J}$, where $\hat \ttt_i$ and $\hat{\boldsymbol\pi}_j$ are from \eqref{eq:cjmle}, and $\ttt_i^*$ and $\boldsymbol\pi_j^*$ are the true parameter values that are required to satisfy the constraints in \eqref{eq:cjmle}. By making use of \eqref{eq:optrate} and by proving extensions of the Davis-Kahan-Wedin sine theorem \citep{davis1970rotation,wedin1972perturbation} for the perturbation of eigenvectors, it can be shown under suitable conditions that
\begin{equation}\label{eq:estitem}
\sqrt{\frac{\sum_{j=1}^J \Vert \hat{\boldsymbol\pi}_j - \boldsymbol\pi_j^* \Vert^2}{J}} = O_p\left(\frac{1}{{\sqrt{N\wedge J}}}\right).
\end{equation}
Under the typical setting where $N$ is much larger than $J$, the rates in \eqref{eq:optrate} and \eqref{eq:estitem} both become $1/\sqrt{J}$; that is,
the accuracy of the JML estimator is mainly determined by the number of items. We  conjecture that  $\max_{j = 1, ..., J} \Vert \hat{\boldsymbol\pi}_j - \boldsymbol\pi_j^* \Vert = o_p(1)$, as both $N$ and $J$ grow to infinity.   This conjecture may be proved by a careful expansion of the joint likelihood function at the constrained JML estimator.

With the above justifications, one may use the JML estimator as an approximation to the MML estimator in large-scale applications, when both $N$ and $J$ are large and the computation of the MML estimator is intensive.

\subsubsection{Full Bayesian estimation.}
Full Bayesian methods have also been developed for the estimation of IRT models. These methods regard the unknown parameters in $\boldsymbol\pi_j$ and $f$ as random variables and impose prior distributions for them. The estimation and associated uncertainty quantification are obtained based on the posterior distributions of the unknown parameters.
MCMC algorithms have been developed for the full Bayesian estimation of different IRT models, especially multidimensional IRT models;
see, for example, \cite{culpepper2015bayesian}, \cite{edwards2010markov} and  \cite{jiang2019gibbs}.

\subsubsection{Other estimation methods. }

We summarize two other estimation methods  commonly used in IRT analysis. One is the conditional maximum likelihood, which essentially follows the ``stochastic subject" regime. This method makes use of a conditional likelihood function that does not contain any person parameters.
This estimator is not flexible enough in that the construction of the conditional likelihood
relies heavily on the raw total score being a sufficient statistic for the person parameter and thus is only applicable to models within the Rasch family \citep{andersen1970asymptotic,andersen1977sufficient,andrich2010sufficiency,rasch1960probabilistic}.

A two-step procedure is typically used to
analyze unidimenisional non-parametric IRT models that assume monotone IRFs  \citep{douglas1997joint,guo2011nonparametric, johnson2006nonparametric}. In the first step, one estimates the individuals' latent trait level based on their total scores. Note that this step only makes sense for unidimensional IRT models with monotone IRFs.
Then in the second step, the IRFs are estimated using a non-parametric regression procedure that regresses the response to each item on the estimated latent trait level. To obtain  consistency in estimating the IRFs, one typically needs the measurement error in the first step to decay to 0, which requires the number of items $J$ to grow to infinity. See \cite{douglas1997joint} for the asymptotic theory.

\subsection{Estimation of Person Parameters}\label{subsec:est_person}

Here we discuss the estimation of person parameters, assuming that item parameters are known. This problem is closely related to the scoring of individuals in psychological and educational measurement. Two settings will be discussed, an offline setting for which response data have already been collected and an online setting for which responses are collected sequentially in real-time.

\subsubsection{Offline estimation.}

We first consider the offline setting for the estimation of person parameters, assuming that
the item parameters are known.
This problem is similar to that of a regression problem.  Specifically,
the likelihood function of person parameters factorizes into a product of the likelihood functions of individual $\ttt_i$s.
As a result, $\ttt_i$ can be estimated by the maximum likelihood estimator, where the likelihood function of $\ttt_i$ is given by
\begin{equation}\label{eq:personlik}
L_i(\ttt_i) = \prod_{j=1}^J g_j(\ttt_i \vert \boldsymbol\pi_j)^{y_{ij}}  (1-g_j(\ttt_i \vert \boldsymbol\pi_j))^{1-y_{ij}}.
\end{equation}
When the distribution $f$ is specified, $\ttt_i$ can also be estimated by a Bayesian method, such as the maximum a posteriori probability (MAP) or expected a posteriori (EAP) estimator. In fact, the posterior distribution of $\theta_i$ is proportional to
$L_i(\ttt_i)f(\ttt_i)$.

The estimation of $\ttt_i$ leads to the
scoring of individuals. Specifically, in a unidimensional IRT model for an educational test,
the estimates $\hat \theta_i$ provide a natural ranking of the test-takers in terms of their proficiency on the construct measured by the test.
Note that
$\boldsymbol\pi_j$ and $f$ may be estimated using the procedures described in Section~\ref{subsec:est_item} above and then treated as known for the estimation of the person parameters. We also point out that some methods described in Section~\ref{subsec:est_item} automatically provide person parameter estimates as byproducts, such as the JML estimator, and the Bayesian estimator based on an MCMC algorithm  in which posterior samples of the person parameters are obtained.


One advantage of scoring by IRT is that linking is automatically performed through the calibration of the items so that the item parameters
are on the same scale as that of the latent traits. As a result, the estimated person parameters are aligned on the same scale, even when different students receiving different subsets of the items. In that case, for each person $i$,
a likelihood function similar to that of \eqref{eq:personlik} can be derived, where items involved in the likelihood function are the ones that person $i$ receives. This property of IRT leads to an important application of IRT, the computerized adaptive testing (CAT). In CAT, a pool of items (also known as an item bank) is pre-calibrated, and test-takers receive different sets of items following a sequential item selection design.
CAT applies an online (i.e., sequential) method for estimating person parameters, which is made possible by techniques from sequential analysis, a branch of statistics concerned with adaptive experimental design and optimal stopping.  We discuss this problem below.


\subsubsection{Online estimation via computerized adaptive testing.} \label{subsec:CAT}


IRT also serves as the foundation for computerized adaptive testing (CAT), a computer-based testing mode under which items are fed to an individual sequentially, adapting to the current knowledge about the individual's latent traits. In educational testing, the use of CAT
avoids giving
capable test-takers too many easy items and giving less capable test-takers  too many difficult ones. Consequently, CAT  can lead to accurate measurement of the individuals with a smaller number of items, in comparison with non-adaptive testing. The concept of adaptive testing was originally conceptualized by \cite{lord1971robbins} in his attempt to apply the stochastic approximation algorithm of \cite{robbins1951stochastic} to design more efficient tests. This idea of adaptive testing was realized in early works on CAT, including \cite{lord1977broad} and \cite{weiss1974strategies,weiss1978proceedings, weiss1982improving}.
A comprehensive review of the practice and statistical theory of CAT can be found in  \cite{chang2015psychometrics}, \cite{van2010elements} and \cite{wainer2000computerized}. 

The CAT problem can be regarded as a sequential experimentation and estimation problem, where an IRT model is assumed with known
item parameters and continuous latent traits.
The aim of CAT is to achieve a pre-specified level of accuracy in estimating each person parameter with a test length as short as possible.
A CAT algorithm has three building blocks, (1) a stopping rule which decides whether to stop testing or not at each step based on the individual's current performance (i.e., performance on the finished items), (2) an item selection rule, which decides which item to give to the test-taker in the next step, and (3) an estimator of the individual's latent traits $\ttt$ and its inference.

 There are statistical problems arising from the above CAT procedure. First, given a stopping rule and an item selection rule, how do we obtain $\hat \ttt$ and its standard error? The setting is different from estimating person parameters based on static item response data, as the responses are no longer conditionally independent given the latent trait level due to sequential item selection. This problem was investigated in \cite{chang2009nonlinear}, where $\hat \ttt$ can be obtained by solving a score equation. Theoretical results of consistency, asymptotic normality, and asymptotic efficiency were established by making use of martingale theory.

Second, how should the item selection rule and stopping rule be designed? This problem is closely related to the early stopping problem and the sequential experimental design problem in the literature of sequential analysis, where the former dates back to  Wald's pioneer works on sequential testing \citep{wald1945sequential,wald1948optimum} and the latter dates back to Chernoff's seminal works on sequential experimental design \citep{chernoff1959sequential,chernoff1972sequential}.
These problems are major topics of sequential analysis; see \cite{lai2001sequential} for a comprehensive review.
More specifically,
the optimal sequential decision on early stopping and item selection
can be formulated under the Markov decision process (MDP) framework \citep{puterman2014markov}, a unified probabilistic framework
for optimal sequential decisions.
In this MDP, the goal is to minimize  a certain loss function (or equivalently, to maximize a certain utility function) that concerns both the accuracy of measurement and the number of items being used, with respect to early stopping and item selection as possible actions that need to be taken at each step based on the currently available information.

A seemingly standard MDP problem, the item selection and early stopping problems in CAT typically cannot simply be solved by dynamic programming,
the standard method for MDPs,
due to the huge state space and action space. In fact, obtaining an exact optimal solution under a nonasymptotic setting is NP-hard under the CAT setting.
Therefore, the developments of CAT procedures are usually guided by asymptotic analysis, heuristics of finite sample performance, and also practical constraints such as speed of the algorithm, item exposure, and balance of contents.
For a practical solution to the CAT problem under unidimensional IRT models,  \cite{lord1980applications} proposed to select the next item
as the one with
the maximum Fisher information at the current point estimate of $\theta$, and \cite{chang1996global} proposed methods based on a Kullback–Leibler (KL) information index, which takes uncertainty in the point estimate of $\theta$ into account under a Bayesian setting. Although these maximum information item selection methods are asymptotically efficient, they often do not perform well at the early stage of a CAT when the estimation of $\theta$ is inaccurate. This is because, an item with maximum information at the estimated $\theta$ may not be informative at the true value of $\theta$ when the true value and its estimate are far apart at the early stage. In addition, the use of maximum information item selection methods could lead to  skewed item exposure rates. That is,
some items could be frequently used in a CAT whereas others might never be used, which may lead to item leakage. To improve item selection at the early stage of a CAT, \cite{chang1999stratified} proposed a multistage item selection method. This method stratifies the item pool into
less discriminating items and discriminating items, where a
discriminating item tends to have a large information index (e.g., Fisher information) at a certain $\theta$ value while a less
discriminating item has a relatively smaller information index
for all values of $\theta$. It uses less discriminating
items early in the test  when estimation is inaccurate, and saves highly discriminating items until later stages.
The stopping of a CAT procedure is often determined by the asymptotic variance of the $\theta$ estimate \citep{weiss1984application} and the sequential confidence interval estimation \citep{chang2005application}.

CAT methods have also been developed under DCMs where the latent variables are discrete. The CAT problem becomes a sequential classification problem under this setting, which is different from sequential estimation. New methods have been developed for the item selection, early stopping, and making   final classification. See  \cite{cheng2009cognitive}, \cite{liu2015rate}, \cite{tatsuoka2003sequential}, and \cite{xu2003simulation}.

The past decade has seen the increasing use of computerized multistage testing \citep{yan2016computerized} in the educational testing industry, a testing mode that can be viewed as a special case of the CAT. Instead of adaptively selecting individual items, computerized multistage testing  divides a test into multiple stages and  adaptively selects a group of items for each stage based on an individual's previous performance. Instead of solving a standard sequential design problem as in CAT, computerized multistage testing involves solving a group sequential design problem. We note that similar group sequential design problems have been widely encountered in clinical trials, for which statistical methods and theory have been developed; see, for example,  Chapter 4, \cite{bartroff2012sequential}.





\subsection{Evaluation of IRT Models}\label{subsec:evaluation}

The psychometric validity of an IRT model relies on the extent to which its assumptions hold.  In what follows, we discuss the evaluation of the overall goodness-of-fit of an IRT model and the assessment of specific model assumptions.

\subsubsection{Overall goodness-of-fit.}

Assessing the overall goodness-of-fit of an IRT model can be cast into testing the null hypothesis of data being generated by the IRT model. In principle, this problem can be solved by Pearson's chi-squared test \citep{pearson1900x} given data with a large sample size. However, as mentioned in Section~\ref{subsec:contingency},
the $J$-way contingency table is typically sparse  when  the number of items $J$ is moderately large, resulting in the failure of the asymptotic theory for Pearson's chi-squared test.

There are two types of methods that give valid statistical inference for sparse contingency tables. The first type is  bootstrap methods \citep[e.g.,][]{collins1993goodness,bartholomew1999goodness}, which is typically time consuming. The second method is based on assessing the fit of
  lower-way marginal tables, such as two- or three-way tables based on item pairs or triplets, rather than a complete table with $2^J$ cells. Asymptotic theory holds for these marginal tables since they have much smaller numbers of cells. Developments in this direction include
    \cite{bartholomew2002goodness},
  \cite{christoffersson1975factor},
\cite{maydeu2005limited,maydeu2006limited}
  and   \cite{reiser1996analysis}.

When an overall goodness-of-fit test suggests a lack of fit, it is desirable to obtain information about which specific assumptions are being violated. 
Methods for assessing individual assumptions of an IRT model will be discussed below. In addition, as statistical models are only an approximation to the real data generation mechanism, IRT models are typically found to lack fit when applying to real-world item response datasets that have a reasonably large sample size \citep{maydeu2006limited}.


\subsubsection{Dimensionality.}\label{subsubsec:3.3.2}

IRT models are often used in a confirmatory manner where the number of latent traits is pre-specified, especially for
unidimensional IRT models. A question that needs to be answered is whether the pre-specified latent dimension is sufficient or some extra dimensions are needed.
It is natural to answer this question by the comparison of IRT models with different numbers of latent traits. For example, to assess unidimensional assumption of the 2PL model, we may compare it with an M2PL model that has two or more latent traits.
While seemly a simple problem of comparing nested models,
the asymptotic reference distribution for the corresponding likelihood ratio test
is not a chi-squared distribution  due to the null model lying at boundary points or singularities of the parameter space in this problem \citep{chen2020note}. This asymptotic reference distribution can be derived using a more general theory for the likelihood ratio test statistic \citep{chernoff1954distribution,drton2009likelihood}.


Alternatively, one may also use an exploratory approach to directly learn the latent dimension from data, and then compare it with the pre-sepecified dimension. See Section~\ref{subsec:structure} for further discussions.

\subsubsection{Local independence.} The local independence assumption plays an essential role in IRT models. It is closely related to the dimensionality assumption that is discussed in Section~\ref{subsubsec:3.3.2}, in the sense that the existence of extra latent dimensions can cause the violation of the local independence assumption. The local independence assumption is often assessed through an analysis of the
residual dependence given the hypothesized latent traits. The goal is to find dependence patterns in data that are not attributable to the primary latent dimensions. Such patterns could reveal how the hypothesized latent structure is violated, and to what extent the violation is.

One type of methods is based on the residuals for lower-way marginal tables \citep[e.g.,][]{chen1997local,liu2013local,yen1984effects}. These methods  search for subsets of items  (e.g., item pairs) that violate the local independence assumption, by defining residual-based diagnostic statistics
and testing the goodness-of-fit of the corresponding lower-way marginal tables based on these statistics. 
Although these methods have reasonably good power according to simulation studies, there is a gap. That is,
the null hypothesis  in the hypothesis tests based on different marginal tables is always ``the fitted IRT model holds for all items", rather than ``local independence holds within the corresponding subset of items". Therefore, from the perspective of hypothesis testing,   when the null hypothesis is rejected for a marginal table, it does not directly lead to the conclusion that the local independence assumption is violated due to the corresponding subset of items.

Alternatively, one can model the local dependence structure. When taking this approach, it is important to impose certain parsimony/sparsity assumptions on the residual dependence structure to ensure the identifiability of the primary dimensions. Several modeling approaches have been taken. First, the local dependence may be modeled by  additional latent traits with a sparse loading structure \citep[e.g.,][]{bradlow1999bayesian,cai2011generalized,gibbons1992full}.
These methods typically require knowing a priori the subsets of items,
which are known as the testlets or item parcels,
that require additional latent traits. 
Second,
copula-based models have been proposed \citep{braeken2011boundary,braeken2007copula}, where copula models are used for
the conditional distribution of  $(Y_1, ..., Y_J)$ given the primary latent dimensions. These models typically impose relatively strong parametric assumptions on the copula function.
Finally, Markov random field IRT models have also been developed \citep{chen2018robust,hoskens1997parametric,ip2002locally,ip2004locally} that model
the conditional distribution of $(Y_1, ..., Y_J)$ given the primary latent dimensions using a Markov random field model.   Markov random field models \citep{koller2009probabilistic,wainwright2008graphical}, also known as undirected graphical models,
are a powerful tool for modeling the dependence structure of multivariate data. Specifically, these models represent the conditional independence relationships between random variables using an undirected graph.
When all the items are binary, this conditional model becomes an Ising model \citep{ising1925beitrag}.  
Specifically,
\cite{chen2018robust}  proposed
a flexible  Markov random field IRT model and developed a Lasso-based method for learning  a sparse  Markov random field.
Comparing with other methods, this approach does not require prior knowledge about the local dependence structure or impose a strong distributional assumption. It also gives a visualization of the local dependence structure, which facilitates the interpretation.

\subsubsection{Item response functions.} Finally, a key assumption of an IRT model is the functional form of the IRFs, especially when parametric IRFs are assumed.
This assumption is commonly assessed by residual analysis.   \cite{hambleton1991fundamentals} first proposed item-specific standardized residuals
for unidimensional IRT models that assess the fit of an IRF at different latent trait levels. \cite{haberman2013generalized} and \cite{haberman2013assessing} modified the method of
\cite{hambleton1991fundamentals} and further developed asymptotic normality results to provide theoretical guarantee for the residual-based statistical inference.


Alternatively, one may also fit IRT models with non-parametric IRFs and then compare the fitted parametric IRFs with their non-parametric counterparts. See Section~\ref{subsec:moreexample} for a review of non-parametric IRT models.

\subsection{Learning the Latent Structure}\label{subsec:structure}

\subsubsection{DIF analysis.}\label{subsubsec:DIF}
DIF analysis, which is introduced in Section~\ref{subsec:moreexample}, is essential for ensuring the validity of educational and psychological measurement. The basic idea of DIF analysis is to compare multiple groups for their performance on one or multiple potential items, taking into consideration that the groups may have different ability distributions. The DIF problem can be viewed as a multi-group comparison problem. However, it differs from standard settings of the analysis of variance or covariance in that the latent trait values, which are not directly observed, need to be controlled in the DIF analysis. Many statistical procedures have been developed for DIF analysis; see \cite{holland2012differential} and \cite{roussos2004differential} for a review.

Consider the DIF formulation described in Section~\ref{subsec:moreexample} under a MIMIC model. Recall that $x_i = 0$ and 1 indicate the reference and focal group memberships, respectively. In the DIF analysis, the IRFs may depend on the group membership. That is, we define a group-specific IRF as $g_j(\ttt\vert \boldsymbol\pi_j^x)  = P(Y_{ij}=1\vert \ttt_i = \ttt, x_i = x)$, where $\boldsymbol\pi_j^0$ and $\boldsymbol\pi_j^1$ denote the group-specific item parameters. In addition, the distribution of $\ttt_i$ can depend on the group membership.  Under this formulation, the DIF items are those satisfying  $\boldsymbol\pi_j^0 \neq \boldsymbol\pi_j^1$. That is, two individuals from different groups would perform differently on a DIF item, even when they have the same latent trait levels. In educational testing, it could mean an item being more difficult for a test taker from one group than one from the other group even though the two test takers have the same ability level, which is a threat to the fairness of the test. Under the path diagram in panel (c) of Figure~\ref{fig:models}, the DIF items are the ones with directed edges from $x$.  The goal of DIF analysis is to detect all the DIF items from data, i.e., to learn the edges from $x$ to $Y_j$s.

The DIF analysis is complicated by the unobserved $\ttt_i$, which needs to be conditioned upon in the group comparison. This complication is due to an intrinsic identifiability issue. More specifically, consider the MIMIC model example given in Section~\ref{subsec:moreexample}.  The model remains unchanged if we subtract any constant $c_0$ from the parameter $\beta$ in the marginal distribution of $\theta_i$ given $x_i$ and further replace $\gamma_j$ in the item response function by $\gamma_j + a_j c_0$. To ensure identifiability, one needs to assume that there are a sufficient number of DIF-free items to fix the latent variable distribution, which becomes assuming $\gamma_j=0$ for many $j$ in this specific MIMIC model.

The Mantel-Haenszel procedure \citep{holland1988differential} is widely adopted for DIF analysis. This procedure compares the two groups' performance on an item based on the Mantel-Haenszel statistic \citep{mantel1959statistical}, using the total score as a matching variable. By stratifying data using the total score, which serves as a proxy of the latent ability being measured, the group difference in the latent ability distributions is accounted for in the comparison. This procedure has a theoretical guarantee under the Rasch model. That is, the $p$-value given by the Mantel-Haenszel procedure is valid when data follow the Rasch model, for which the total score is a sufficient statistic for the person-specific latent variable. Note that the Mantel-Haenszel procedure implicitly assumes that the rest of the items are DIF-free to fix the latent variable distribution when analysing one item.

Another popular approach to DIF analysis takes a general nonparametric IRT setting, as first proposed in \cite{shealy1993model,shealy1993item}. 
Specifically, \cite{shealy1993model,shealy1993item} laid a theoretical foundation for DIF analysis from a nonparametric multidimensional IRT perspective. Under this general statistical framework, they provided mathematical characterizations of different DIF types and further pointed out the importance of simultaneously analyzing the bias in a set of items. They also proposed a nonparametric procedure, known as the SIBTEST. Similar to the Mantel-Haenszel procedure, SIBTEST compares the overall performance of the reference and focal groups on one or multiple  suspected items to detect DIF, by matching individuals on a known subset of DIF-free items. Further developments in this direction include \cite{chang1996detecting}, \cite{douglas1996kernel}, \cite{jiang1998improved}, \cite{li1996new}, \cite{nandakumar2004evaluation}, \cite{stout1997multisib},   among others.

Both the Mantel-Haenszel and SUBTEST procedures are hypothesis-testing-based. Possibly more naturally, one can view the DIF problem as a model selection problem. More specifically, in the specific MIMIC example discussed previously, the DIF problem becomes to find the non-zero $\gamma_j$s, under the sparsity assumption described above. Taking this view, several regularized estimation methods have been developed for the detection of DIF items under MIMIC-type models; see \cite{belzak2020improving}, \cite{huang2018penalized}, \cite{magis2015detection}, and \cite{tutz2015penalty}. In these methods, the DIF-free items are identified by LASSO-type regularizations.

\subsubsection{Exploratory analysis of latent structure.}\label{subsubsec:Q}

As pointed out in Sections~\ref{subsec:model} and \ref{subsec:moreexample}, the $Q$-matrix structure plays an important role in multidimensional IRT models.
Although domain experts usually have ideas about the latent traits underlying the items, it is also helpful to consider an exploratory analysis setting under which the meanings of the latent traits and the structure of the $Q$-matrix are unknown. Under this setting, we may learn the latent traits and the $Q$-matrix from data and use the learned structure to validate the current hypotheses  or generate new hypotheses about data.

Analytic rotation methods, which are originally developed
for exploratory factor analysis, can be used to solve this structural learning problem under several multidimensional IRT models, including the M2PL model described in Section~\ref{subsec:model} and its probit counterpart in Section~\ref{subsec:FA}. More specifically, consider an exploratory M2PL model, for which there are no zero constraints on the loading matrix $A$, and  $\ttt_i$ follows a
standard multivariate normal distribution.
Let $\hat A$ be an MML estimate of the loading matrix $A$, which is not unique due to the  rotational indeterminacy. More specifically, replacing $\hat A$ by $\hat A T$ and replacing the covariance matrix of $\ttt_i$ by $T^{-1}(T^{-1})^\top$ lead to the same marginal likelihood function value.
Matrix $T$ is called an oblique rotation matrix when the diagonal entries of $T^{-1}(T^{-1})^\top$ all take value one.
In the special case when the matrix $T$ is an orthonormal matrix, $T^{-1}(T^{-1})^\top$ becomes an identity matrix and the matrix $T$ is known as an orthogonal rotation matrix. A rotation method minimizes a loss function $H(\hat A T)$ with respect to an oblique or
orthogonal rotation matrix $T$, and then uses $\hat A \hat T$ as the final estimate of the loading matrix where $\hat T$ denotes the corresponding minimizer. The loss function is designed to impose approximate sparsity on the resulting loading matrix $\hat A \hat T$,  implicitly assuming that there is an unobserved sparse $Q$-matrix.
Different analytic rotation methods have been proposed that consider different loss functions, including \cite{jennrich2004rotation,jennrich2006rotation},  \cite{jennrich2011exploratory}, \cite{jennrich1966rotation},  \cite{kaiser1958varimax}, among others. A comprehensive review of rotation methods can be found in \cite{browne2001overview}.  Among these methods, we draw attention to a special case of the component loss
functions proposed in \cite{jennrich2004rotation,jennrich2006rotation}. This loss function takes the form $H(A) = \Vert A\Vert_1 = \sum_{j=1}^J\sum_{k=1}^K \vert a_{jk} \vert  $. As we discuss below, it is closely related to the regularized estimation approach to this problem.
The analytic rotation methods have some limitations. First, these methods do not give a sparse estimate of the loading matrix, though they may resolve the rotational indeterminacy issue and lead to a consistent estimator of the true loading matrix \citep{jennrich2004rotation,jennrich2006rotation}.
Consequently, obtaining the $Q$-matrix estimate and further, the interpretation of the latent traits are not straightforward. Second, these methods are only applicable to multidimensional IRT models in which the latent traits follow a multivariate normal distribution and the IRFs take a simple generalized linear model form.

Regularized estimation methods have also been proposed
that do not suffer from the above limitations. Specifically, consider the M2PL model. A LASSO-based regularization method solves the optimization
\begin{equation}\label{eq:regmle2}
\begin{aligned}
(\hat{\boldsymbol\pi}_1^{\lambda}, ..., \hat{\boldsymbol\pi}_J^{\lambda}, \hat \Sigma^{\lambda}) = & \argmin_{\boldsymbol\pi_1, ..., \boldsymbol\pi_J, \Sigma} \left(- \log L_{ML} (\boldsymbol\pi_1, ..., \boldsymbol\pi_J, \Sigma) + \lambda\sum_{j=1}^J \sum_{k=1}^{K} \vert a_{jk}\vert\right),\\
\mbox{s.t. } &  \sigma_{kk} = 1, k=1, ..., K.
\end{aligned}
\end{equation}
With slight abuse of notation, $L_{ML} (\boldsymbol\pi_1, ..., \boldsymbol\pi_J, \Sigma)$ denotes the marginal likelihood function. Note that we assume $\ttt_i$
to follow a multivariate normal distribution with mean zero and covariance matrix $\Sigma$ whose diagonal entries all take value one. The Lasso penalty can lead to a sparse estimate of the loading matrix $A$, which further gives an estimate of the $Q$-matrix. In \eqref{eq:regmle2}, $\lambda$ is a non-negative tuning parameter that can be chosen by information criteria or cross-validation. Note that this approach is closely related to the rotation approach. That is, when $\lambda$ converges to zero, the solution in \eqref{eq:regmle2} will converge to the oblique rotation solution under the $L_1$ loss function mentioned above. We note that the LASSO penalty \citep{tibshirani1996regression} in \eqref{eq:regmle2} can be replaced by non-convex penalty functions, including the widely used SCAD and MCP penalties \citep{fan2001variable,zhang2010nearly}. Compared with the analytic rotation methods, this approach is more general and can be applied to a wide range of IRT models, including multidimensional IRT models with continuous latent traits \citep{sun2016latent} and DCMs \citep{chen2015statistical, xu2018identifying}.
The optimization for these regularized estimators, such as \eqref{eq:regmle2}, is more complicated than that for MML estimation when the latent dimension is high due to the non-smooth penalty term. To solve this optimization problem, \cite{zhang2020computation} proposed a stochastic approximation algorithm that combines a standard stochastic approximation algorithm with the proximal method \citep{parikh2014proximal}  to handle non-smooth penalty terms.

Besides the above methods, several alternative methods have been developed for DCMs. In particular, \cite{chiu2013statistical} and \cite{de2016general} proposed iterative search algorithms for $Q$-matrix validation and refinement.
\cite{chen2018bayesian} and \cite{culpepper2019estimating}
proposed full Bayes methods for $Q$-matrix learning.



\subsubsection{Learning the latent dimension.} %
Statistical methods have been developed for determining the latent dimension $K$ under an exploratory analysis setting. Some of them only apply to the probit model described in Section~\ref{subsubsec:uv}. These methods arise from exploratory linear factor analysis, by making use of the connection between the probit model and the linear factor models. More precisely, consider the underlying variable formulation for the probit model given in Section~\ref{subsubsec:uv}.
When the true latent dimension is $K$, then the true correlation matrix of the underlying variables can be decomposed as the sum of a rank-$K$ matrix and a diagonal matrix. Consequently, the latent dimensionality can be determined based on the estimated correlation matrix, which, as discussed in Section~\ref{subsubsec:lie}, can be efficiently computed. These methods include
eigenvalue thresholding \cite{kaiser1960application}, subjective search for eigengap based on scree plot
\cite{cattell1966scree}, parallel analysis \citep{horn1965rationale}, among others.

Besides, information criteria are commonly used for determining the latent dimension.
Specifically, the Akaike information criterion \citep{akaike1974new} and Bayesian information criterion \citep{schwarz1978estimating}, which are calculated based on the marginal likelihood, are widely used for comparing IRT models of different latent dimensions \citep{cohen2016information}. Unlike the  exploratory-factor-analysis-based methods, these information criteria are applicable to all parametric IRT models that treat the latent variables as random. Treating the latent variables as fixed parameters, \cite{chen2020determining} proposed an information criterion based on the joint likelihood. Their method guarantees consistent selection of the latent dimension under a high-dimensional setting where   the sample  $N$ and the number of items $J$ grow to infinity simultaneously, and the latent dimension may also diverge.

Under nonparametric IRT modelling,  \cite{stout1987nonparametric,strout1990new} introduced the concept of essential dimensionality, aiming at finding the dominant dimensions in the data while allowing for the presence of ignorable nuisance dimensions. 
Several methods for assessing and choosing the essential dimension were proposed; see \cite{stout1987nonparametric}, \cite{strout1990new}, \cite{nandakumar1993refinements}, and \cite{zhang1999theoretical}.  They established consistency results
under a double asysmptotic regime where both $N$ and $J$ grow to infinity.

\section{New Challenges and Opportunities} \label{sec:future}

%
%
%

\subsection{Measurement in Big Data Era}
In the previous discussions, the IRT framework is tailored to the analysis of nicely structured item response data from traditional educational and psychological assessments. 
This specialization is a limitation for the generalization and extension of IRT methods to measurement problems based on more complex human behavioral data, which are becoming increasingly
 prevalent in the big data era.

For example, with the wide use in daily teaching of digital devices and information technology (e.g., tablets, computers, and online grading systems), students' item responses to all the homework/exam questions throughout their entire academic life can be recorded. In addition,
students' learning process data may also become available, including the processes of reading, searching, and problem-solving \citep[e.g.,][]{zhang2016smart}.
Moreover, new measurement tools are being developed to measure people's skills that are difficult to assess by traditional paper-and-pencil tests, such as critical thinking, collaboration, and complex problem-solving \citep{organisation2018future}. For example, complex problem-solving skills are currently assessed in large-scale assessments, including the  PISA, Programme for the International Assessment of Adult Competencies (PIAAC) and the National Assessment of Education Progress (NAEP)
using simulated tasks in computers. Data collected were not only test-takers’ success or failure in solving such tasks, but also their entire problem-solving processes recorded by computer log files that may contain important information about their complex problem-solving ability. In psychology, researchers tend to study psychological constructs using not only item response data but also a much wider range of daily behavioral data including those from social media, facial expression, and writing and speech \citep[e.g.,][]{inkster2016decade,nave2018musical,wang2018deep}.

As revealed by the above examples, human behavioral data of much higher volume, variety, and velocity are being collected nowadays.
It is widely believed that rich information in such big data can substantially improve our understanding of human behavior and facilitate its prediction. The obtained insights and prediction, assisted by information technology, can lead to effective personalized interventions, such as personalized learning systems that suggest learning strategies adaptive to person-specific learning history and characteristics and personalized monitoring and intervention system for mental health patients. 


On the one hand, such big data can still be viewed as people's responses to complex items in our study and daily life. Thus, the fundamental latent dimension view of IRT should still apply to
such big data. It is reasonable to believe that people's behaviors are mainly driven by some intrinsic and relatively stable latent traits. Therefore, such latent traits may be measured and be used to predict people's future behaviors.
On the other hand,
compared with traditional item response data, big data have a far more complex structure and a much higher volume. As a result, most of the traditional IRT models and algorithms are no longer suitable.

The question now becomes how the methodological framework of IRT may be extended to embrace the opportunities and challenges in the big data era. While taking  educational and psychological theories as the theoretical ground and ensuring the primacy of measurement validity,
more flexible models and computationally faster algorithms from statistical and machine learning fields may be incorporated into the IRT framework to better measure human traits. The following sections will discuss two future directions that we believe are important and require methodological developments. We also note that many other promising directions are omitted here due to the space limit.



\subsection{Learning the Structure of Fine-grained Constructs}

The ultimate goal of psychological measurement is to understand human behavioral patterns, i.e., to obtain knowledge about how people learn, think, and make decisions. Such knowledge is often gained from taxonomic analysis of individual differences. With big data,
evidence-based taxonomy of educational or psychological constructs at a higher resolution becomes possible.
For example, in the field of psychopathology, the traditional classification system of mental health disorders, widely adopted in the standard diagnostic criteria (\textit{Diagnostic and Statistical
Manual of Mental Disorders} \citep[DSM;][]{american2013diagnostic} and the International Classification of Diseases \citep[ICD;][]{world1992icd}), has received criticism for having limited reliability and validity  \citep{kotov2017hierarchical}.
A better-structured, finer-grained, and evidence-based taxonomic system needs to be developed \citep{kotov2017hierarchical}.

For another example, knowledge maps play an important role in designing a curriculum or in an e-learning system for conceptualizing and organizing the structure of the basic units of learning. They can be a powerful tool for guiding teaching and learning \citep[][Chapter 4]{knight2013high}.  Each node of a knowledge map represents a concept or skill. An edge between two nodes captures their relationship, such as whether acquiring one concept requires another as a prerequisite.
Examples can be found in \cite{kingston2017use} on the development and use of digital knowledge maps for supporting formative assessments in mathematics, where a knowledge map can consist of
thousands of nodes (points of mathematical knowledge)
and thousands of connections between these nodes. Traditionally, knowledge maps were designed by domain experts, which are sometimes  subjective. Rich data from daily teaching and e-learning platforms can be used to facilitate the design of knowledge maps. Data can tell us whether a node needs to be split into multiple ones and whether learning a certain concept/skill relies on certain concepts/skills as prerequisites.

These taxonomic
problems are closely related to, but more complex than,
the $Q$-matrix learning problem reviewed in Section~\ref{subsubsec:Q}. Specifically, data are more complex, multi-modal, and collected sequentially.  For example, to design a knowledge map, one needs to consider the existing curriculum design,  data from domain experts on each node's definition and the relationships between nodes, and students' learning history data.
In addition, the improvement of the knowledge map is sequential, iterating between the steps of data collection under an existing knowledge map and updating the knowledge map. Moreover,   a knowledge graph takes the form of a directed graph \citep{kingston2017use}, and the taxonomic structure of psychopathology is believed to be efficiently captured by a tree structure with many levels \citep{kotov2017hierarchical}. These structure learning problems are harder to formulate and solve than the previously discussed $Q$-matrix problem.
Similar to the learning of $Q$-matrix in multidimensional IRT models \citep{chen2015statistical,liu2012data,liu2013theory, xu2018identifying},
these taxonomic problems may be formulated under latent variable models and solved by similar regularized estimation methods for structure learning.

\subsection{Measurement from a Prediction Perspective}\label{subsec:prediction}

Measurement and prediction are closely related.
A major purpose of measurement is to make predictions, even for the traditional use of measurement results.
For instance, an important use of test scores in education is to find the strengths and weaknesses of a student, which can provide predictions of the potential learning outcome in their subsequent learning trajectory. Accurate forecasting of students' potential learning outcomes can help teachers to make a tailored teaching plan. In mental health, an individual's responses to a mental health questionnaire not only tell clinicians the individual's current mental health status but also reveal the likelihood of developing certain mental health disorders in the near future.

A related and more challenging problem
is to make many predictions simultaneously.
One such application is in personalized learning, where personalized formative assessment questions need to be selected from a large pool of items to meet the need of every student. A good formative assessment should achieve a balance in both its contents and difficulty/complexity, so that best serves the purpose of monitoring student learning. In this problem, to assemble a suitable set of formative assessment items, we need to predict every student's performance on each of the candidate items in the pool,  which involves hundreds or even more simultaneous predictions  for each student.
Data in these predictive analyses may not be well-organized and may have massive missingness compared to conventional IRT analysis.

Latent variable models are a powerful tool that supports making such predictions because these models can learn the joint distribution of multivariate data even when massive missing data exist. One or multiple predictions can be made for the unobserved data given the observed ones with the learned joint distribution. Moreover,  assumptions of latent variable models essentially reduce the dimensionality of the joint distribution of multivariate data, which will reduce the variance of prediction according to basic principles of statistical learning. Note that latent variable models, including nonlinear latent factor models and neural network models, have been popular tools for prediction tasks in other areas. These models may be borrowed to solve prediction problems in educational and psychological measurement. We note that some new developments have been made in this direction for
 analyzing and predicting
log-file data of complex problem-solving processes; see \cite{chen2020continuous}, \cite{tang2020latent}, \cite{tang2021exploratory}, \cite{wang2020subtask}, and \cite{zhang2021accurate}.

Conventional IRT analysis takes a causal explanation perspective, which has substantial distinctions from the prediction perspective.
While a model with high explanatory power may also have high predictive power, it is not always the case. It is well-known that a model selected for causal explanation may not have the best prediction performance \citep{shmueli2010explain}.
Therefore, a predictive model may take a very similar form as a multidimensional IRT model, estimating and selecting a predictive model often need to follow a different criterion from conventional IRT analysis.
New criteria for estimation and model selection, such as cross-validation, need to be developed.

Learning a predictive model from data is computationally intensive, involving an iterative process that alternates between model fitting (solving an optimization problem) and model evaluation (via cross-validation). The computation becomes more demanding when many latent variables are needed to capture an individual's behavioral patterns. The problem is further exacerbated when online predictions need to be made (e.g., personalized learning). Given such high computational demand, it
may be more suitable to use joint-likelihood-based estimators
that treats the person-specific latent variables as fixed parameters \citep{chen2018joint, chen2019structured} for parameter estimation,
due to its computational advantage.




Moreover, in some human behavior prediction problems, such as personalized learning, an individual's latent characteristics can substantially change within a relatively short period of time.
Dynamic latent variable models need to be developed for solving such problems. We note that there have been some works in this direction, including \cite{chen2018hidden} and \cite{wang2018tracking}.

Finally, we point out that fairness is a key issue when making predictions, especially when interventions (e.g., learning interventions) are made based on the prediction results. The fairness issue is closely related to the DIF problem discussed in Section~\ref{subsubsec:DIF} that has also received wide attention in the statistics and machine learning communities \citep[e.g.,][]{barocas-hardt-narayanan}. Further developments that extend DIF analysis to ensure fairness are needed in the predictive modeling of multivariate behavioral data.

\section{Discussions}\label{sec:dis}


In this paper, we provide an overview of the statistical framework of IRT, and further showcase several applications in classical educational/psychological testing, such as test scoring, item calibration, computerized adaptive testing, differential item functioning, among others. Note that this list of topics is far from complete due to the space limit. Several important applications of IRT are not reviewed, including the detection of fraudulent behaviors and test assembly, among others. As we can see from the discussion, latent variables play a central role in IRT, serving as its philosophical foundation and methodological underpinning. We make the connection between the latent factor framework of IRT and new statistical techniques such as nonparametric methods, matrix completion, regularized estimation, and sequential analysis, and further offer a view of psychological measurement from the prediction perspective.




Thanks to the development of technology and data science, a personalization revolution is ongoing in all areas of our life, including mental health, education, social networking, among others \citep{lee2018ai}. As a scientific discipline studying the measurement of individual characteristics supported by statistical methodology and substantive theory from psychology and education, psychometrics will play a key role in this personalization revolution. It will contribute to a better society, for example, by suggesting better learning strategies for learners and by providing early warnings to mental health patients through more accurate measurement and measurement supported prediction \citep{kapczinski2019personalized,zhang2016smart}. As previously discussed, these problems need to be solved from two different perspectives -- high-resolution taxonomic analysis for better explaining individual differences and predictive analytics for making personalized recommendations. Latent variable models and the previously discussed computationally efficient methods for the structure learning and estimation of latent variable models may play an important role in solving these new problems.

\bibliographystyle{apalike}
\bibliography{ref}

\begin{thebibliography}{}
\bibitem[Pearson, 1900b]{pearson1900x}
Pearson, K. (1900b).
\newblock On the criterion that a given system of deviations from the probable
  in the case of a correlated system of variables is such that it can be
  reasonably supposed to have arisen from random sampling.
\newblock {\em Philosophical Magazine}, 50:157--175.
\end{thebibliography}

\end{document}